\documentclass[a4paper,twocolumn,11pt,accepted=2024-11-12]{quantumarticle}
\pdfoutput=1
\usepackage[utf8]{inputenc}
\usepackage[english]{babel}
\usepackage[T1]{fontenc}
\PassOptionsToPackage{compress}{natbib}
\usepackage[numbers]{natbib}
\usepackage{listings}
\usepackage{braket}
\usepackage{amsmath}
\usepackage{amssymb}
\usepackage{hyperref}
\usepackage[resetlabels]{multibib}
\setcitestyle{numbers}

\begin{document}

\title{Problem-tailored Simulation of Energy Transport on Noisy Quantum Computers}

\author{I-Chi Chen}
\email{ichen@iastate.edu}
\thanks{I.-C.C. and K.P. have made equal contribution to this work.}
\affiliation{Department of Physics and Astronomy, Iowa State University, Ames, Iowa 50011, USA}
\affiliation{Ames National Laboratory, Ames, Iowa 50011, USA}

\author{Kl\'ee Pollock}
\email{kleep@iastate.edu}

\affiliation{Department of Physics and Astronomy, Iowa State University, Ames, Iowa 50011, USA}

\author{Yong-Xin Yao}
\email{ykent@iastate.edu}
\affiliation{Department of Physics and Astronomy, Iowa State University, Ames, Iowa 50011, USA}
\affiliation{Ames National Laboratory, Ames, Iowa 50011, USA}

\author{Peter P.~Orth}
\email{peter.orth@uni-saarland.de}
\affiliation{Department of Physics and Astronomy, Iowa State University, Ames, Iowa 50011, USA}
\affiliation{Ames National Laboratory, Ames, Iowa 50011, USA}
\affiliation{Department of Physics, Saarland University, 66123 Saarbr\"ucken, Germany}

\author{Thomas Iadecola}
\email{iadecola@iastate.edu}
\affiliation{Department of Physics and Astronomy, Iowa State University, Ames, Iowa 50011, USA}
\affiliation{Ames National Laboratory, Ames, Iowa 50011, USA}

\begin{abstract}

The transport of conserved quantities like spin and charge is fundamental to characterizing the behavior of quantum many-body systems.
Numerically simulating such dynamics is generically challenging, which motivates the consideration of quantum computing strategies.
However, the relatively high gate errors and limited coherence times of today's quantum computers pose their own challenge, highlighting the need to be frugal with quantum resources.
In this work we report simulations on quantum hardware of infinite-temperature energy transport in the mixed-field Ising chain, a paradigmatic many-body system that can exhibit a range of transport behaviors at intermediate times.
We consider a chain with $L=12$ sites and find results broadly consistent with those from ideal circuit simulators over 90 Trotter steps, containing up to 990 entangling gates.
To obtain these results, we use two key problem-tailored insights.
First, we identify a convenient basis---the Pauli-$Y$ basis---in which to sample the infinite-temperature trace and provide theoretical and numerical justifications for its efficiency relative to, e.g., the computational basis.
Second, in addition to a variety of problem-agnostic error mitigation strategies, we employ a renormalization strategy that compensates for global nonconservation of energy due to device noise. 
We discuss the applicability of the proposed sampling approach beyond the mixed-field Ising chain and formulate a variational method to search for a sampling basis with small sample-to-sample fluctuations for an arbitrary Hamiltonian.
This opens the door to applying these techniques in more general models.
\end{abstract}

\maketitle

\section{Introduction}
Developing an understanding of transport in interacting quantum many-body systems is a topic of substantial renewed interest. Already in one-dimensional (1D) lattice spin models there is a rich array of different potential behaviors for the late-time transport of conserved quantities such as spin and energy. Both phenomenological methods applicable to integrable systems~\cite{Doyon20} as well as numerical simulation methods such as tensor networks (TNs)~\cite{Schollwock11,Orus2014} have led to a number of tentative conclusions about 1D transport. On one hand, it is expected that diffusion is the generic behavior of interacting systems with few conservation laws~\cite{Bertini2021}. On the other hand, strongly disordered systems may display subdiffusive or localized dynamics~\cite{Znidaric16,Basko06,Gornyi-PRL-2005, Yu16, schulzEnergyTransportDisordered2018, doggenManybodyLocalizationLarge2021} as do systems with multipole conservation laws~\cite{feldmeier_anomalous_2020}. Spin transport in the 1D Heisenberg model is superdiffusive~\cite{Wei22,Keenan22,Gopalakrishnan23} and, remarkably, energy transport in kinetically constrained models can potentially be superdiffusive as well~\cite{Ljubotina23}.

In systems far from integrability, conclusively establishing the nature of transport with classical numerical approaches via exact dynamical simulations is extremely challenging.
To do so, one must simulate the dynamics of large systems out to late times where all transients of a microscopic origin have disappeared and a hydrodynamic description of densities corresponding to conserved quantities is expected to apply. At the same time, exact numerical approaches are limited to small system sizes, while TN methods may be limited to early times or exhibit subtle systematics~\cite{Haegeman11,Haegeman16,Leviatan17,White18,Kloss18,Rakovszky22,Yoo23}.
A natural question, therefore, is whether quantum computers can help. 
Indeed, a few works have already proposed quantum algorithms for this problem, including an algorithm for simulating spin transport using random quantum circuits~\cite{Richter21,Keenan22}.

The main challenge in quantum simulation is errors due to environmental decoherence or imperfect realizations of quantum operations.
The cumulative effect of these errors limits the depth of quantum circuits that can be faithfully executed on hardware. 
In the interest of demonstrating non-trivial quantum simulations while still in the era of noisy intermediate scale quantum (NISQ) computing~\cite{Preskill18}, a variety of error suppression and mitigation techniques have been developed to manage these limitations. 
At the hardware level, one can apply sequences of dynamical decoupling pulses~\cite{Lorenza98, Pokharel18, Jurcevic2021} on the idle qubits to
extend their relaxation and coherence times.
One can also create custom pulse-level gates to reduce circuit duration~\cite{Stenger2021,Kim21}. 
Quantum error mitigation on the algorithmic level~\cite{caiQuantumErrorMitigation2022b} such as zero noise extrapolation (ZNE) ~\cite{Li17,Temme17,Kandala19,Giurgica-Tiron-ZNE-2020, berthusenQuantumDynamicsSimulations2022a, kimEvidenceUtilityQuantum2023} and probabilistic error cancellation (PEC)~\cite{Temme17,Suguru18, mcdonoughAutomatedQuantumError2022, van_den_Berg21} are widely used for quantum simulations on hardware.

In this work we focus on quantum simulation of energy transport in the 1D mixed-field Ising model (MFIM), a prototypical example of a nonintegrable quantum many-body system. We find that standard methods like ZNE and dynamical decoupling are insufficient to capture the correct dynamics of a Trotterized MFIM evolution at system size $L=12$. However, by employing a renormalization strategy that follows naturally from the method we use to sample correlation functions, we gain an empirical factor of about $5$ in the time out to which the dynamics can be correctly captured (corresponding to $90$ total Trotter steps). For an $L=12$ chain, this is sufficient to extract a power-law scaling exponent characterizing the energy transport (on the accessible time scales) that is consistent with classical simulations. 

Our sampling method approximates unequal-time energy density correlators at infinite temperature using random $y$-basis product states (i.e., product eigenstates of single-site Pauli $Y$ operators). For the MFIM, we show with both numerics and analytics that this sampling method has only an $O(1)$ and exceptionally small sampling complexity, making it suitable for NISQ devices. At the same time, to measure unequal-time correlators on hardware, we employ an ancilla-free protocol requiring only direct measurements~\cite{Mitarai19}, which leads to $16$ distinct circuit evaluations at each Trotter step (independent of $L$). Finally, only local operators are measured in this protocol, leading to an $L$-independent shot noise. For a fixed Trotter step size, the algorithm therefore comes with an $O(t)$ quantum circuit complexity, where $t$ is the total evolution time.   

The remainder of the paper is organized as follows. In Sec.~\ref{Sec:Rydberg}, we define the energy density operators for the MFIM along with various quantities that are generally used to diagnose the transport behavior. In Sec.~\ref{Sec:Y-basis}, we discuss the $y$-basis sampling method and point to Appendix~\ref{sec:justify_y}, which contains further numerical and analytical arguments for the effectiveness of sampling in the $y$-basis. In Sec.~\ref{Sec:QC}, we explain our protocol for measuring $y$-basis sampled correlation functions on a quantum computer, discuss the problem-tailored renormalization-based error mitigation strategy, and present results calculated on an IBM quantum processing unit (QPU). 
In Sec.~\ref{sec:generalize} we provide some steps towards generalizing the sampling approach beyond the MFIM.
In particular we formulate a variational approach to search for a sampling basis with small sample-to-sample fluctuations for an arbitrary Hamiltonian.
Notably, this variational method allows for more general sampling bases to be explored, including ones with nonzero entanglement.
Finally, Sec.~\ref{Sec:conclude} concludes and offers an outlook.

\section{Energy transport}\label{Sec:Rydberg}

Motivated by experimental realizations of the MFIM in Rydberg atom arrays~\cite{Bernien17,Bluvstein21}, we focus on the following formulation of the MFIM:
\begin{align}
\label{eq:HRydberg}
H = 4V\sum^{L-1}_{i=1} n_in_{i+1}+\Omega\sum^L_{i=1} X_i,
\end{align}
where $n_i=\frac{I+Z_i}{2}$ and $Z_i (X_i)$ is the Pauli $Z(X)$ operator acting on site $i$.
In Rydberg atom arrays, the transverse field $\Omega$ is set by the Rabi frequency of a laser that pumps the atoms between their ground and Rydberg states with $\braket{Z_i}=+1$ and $-1$, respectively.
Van der Waals interactions between excited Rydberg atoms generate a long-range density-density interaction whose scale is set by the nearest-neighbor interaction strength $V$. We neglect longer-range couplings (which fall off as $\sim|i-j|^{-6}$) in the present work as they do not qualitatively affect the energy transport.
Rewritten in terms of Pauli operators, Eq.~\eqref{eq:HRydberg} becomes 
\begin{equation}\label{eq:H}
    H = N\sum_{i=1}^{L} h_i,
\end{equation}
with $h_i = N^{-1}[\Omega X_{i}\!+\!2VZ_{i}\!+\!\frac{V}{2}\left(Z_{i}Z_{i+1}\!+\!Z_{i-1}Z_{i}\right)]$ for $i$ in the bulk of the chain and 
\begin{align}
\label{eq:hi}
h_{i}
\!=\!
\frac{1}{N}
\!
\begin{cases}
\Omega X_{1}+VZ_{1}+\frac{V}{2}Z_{1}Z_{2} \ &i=1\\
\Omega X_{L}+VZ_{L}+\frac{V}{2}Z_{L-1}Z_{L} &i=L
\end{cases}
\end{align}
at the edges. Here, $N=(\Omega^{2}+\frac{9}{2}V^{2})^{1/2}$ is a normalization constant chosen so that $\braket{h_i^2}_\infty=1$ for $i$ in the bulk of the chain, where $\langle A \rangle_\infty = \text{tr}(A)/d$ with $d=2^L$ is the infinite-temperature average of an observable $A$. The infinite temperature energy transport can be characterized via the real-valued correlation functions
\begin{align}
\label{eq:corr}
    C_{ij}\left(t\right) = \langle h_i(t)h_j(0) \rangle_\infty
\end{align}
where $B(t) = e^{iHt} B e^{-iHt}$ for an observable $B$. Eq.~\eqref{eq:corr} describes the spatial energy distribution upon evolving the slightly out of equilibrium initial state $\rho_\epsilon = e^{\epsilon h_j}/\text{tr}(e^{\epsilon h_j})$ representing the injection of a small amount of energy density on site $j$.
For small $\epsilon$,
\begin{equation}
    \text{tr}( \rho_\epsilon(t) h_{i} ) = \epsilon \braket{h_i(t) h_j(0)}_\infty + O(\epsilon^2);
\end{equation}
hence, $C_{ij}(t)$ are linear response coefficients. Unless otherwise specified, we fix $j=L/2$ and vary $i=L/2+r$ as a function of $r$, defining $C_r(t) \equiv C_{ij}(t)$. 

First taking the limit of $L\rightarrow \infty$, the $C_r(t)$ are expected to admit a hydrodynamic description at long times, 
such that they decay as $C_r\left(t\right)\sim t^{-1/z}$ where $z$ is the dynamical exponent classifying the transport \cite{Bertini2021}. 
When $z>2$ the energy transport is subdiffusive, and when $z<2$ it is superdiffusive. In particular for ballistic transport, which occurs in the presence of a global conserved current, one has $z=1$. For a non-integrable model one generally expects diffusion, corresponding to $z=2$, but other types of transport are also possible as discussed in the introduction.

A related quantity that can be used to diagnose transport behavior is the spatial variance (SV) of the quasi-probability distribution 
over $r$ given by $\tilde{C}_r(t) = C_r(t)/C$, where
 \begin{equation}
\label{eq:const}
   C = \sum_{r=-L/2+1}^{L/2}C_{r}\left(t\right)=\frac{\Omega^{2}+5V^{2}}{\Omega^{2}+\frac{9}{2}V^{2}}
\end{equation}
is a normalization factor introduced so that $\sum_r \tilde C_r(t) = 1$. The fact that $C$ is a time independent constant follows from energy conservation and will later be a crucial ingredient in simulating the transport on a noisy quantum device. The spatial variance (SV) is then
\begin{equation}\label{eq:SV}
  \tilde{\Sigma}^{2}(t)=\sum_{r}r^{2}\tilde{C}_{r}(t) - \bigg(\sum_{r}r \tilde{C}_{r}(t)\bigg)^2
\end{equation}
and is expected to grow as $\tilde{\Sigma}^{2}\left(t\right)\sim t^{2/z}$.

For generic model parameters, the MFIM is nonintegrable with no conserved quantities except energy, and we thus expect it to display diffusive energy transport, i.e.~with a late-time dynamical exponent $z=2$. However, the model~\eqref{eq:HRydberg} also has a natural parameter $\Omega/V$ that can tune the apparent transport behavior at intermediate times. In the limit $\Omega/V\rightarrow 0$, the model maps to the ``PXP model"~\cite{Fendley04,Turner18a,Turner18b}, where pairs of sites for which $\braket{n_in_{i+1}}=1$ effectively become frozen. The prevalence of such configurations leads to exponentially suppressed infinite-temperature transport. Conversely, in the limit $\Omega/V\rightarrow \infty$ the model exhibits ballistic transport in the prethermal regime as the $x$-basis magnetization becomes a conserved quantity and spin flips in that basis become well-defined quasiparticles. In our IBM hardware experiments, we simulate $\Omega/V = 2,3,6$ looking for signatures of effective diffusive and superdiffusive behavior at the accessible timescales.

\section{Sampling Methods}\label{Sec:Y-basis}

\subsection{General approaches}

Let $Q= h_i(t)h_j(0)$, such that $C_{ij}(t)\equiv\braket{Q}_\infty$. A direct and exact approach to computing an infinite-temperature correlation function such as $\braket{Q}_\infty$ would be to pick a complete basis of states $\{\ket{\varphi_k}\}_k$, compute all $\braket{\varphi_k|Q|\varphi_k}$, and then average the results over $k$. A convenient sampling basis could be, e.g., product states, which are cheap to produce on a QPU and to store on a classical computer. By choosing $\ket{\varphi_k}$ to be random product states, one could instead imagine Monte-Carlo sampling $\braket{Q}_\infty$, incurring a statistical error proportional to $S^{-1/2}$ where $S$ is the number of samples. A problem-tailored method of this type has been used for spin transport: Ref.~\cite{joshi_observing_2022} uses random pairs of $z$-basis product states to sample spin-spin correlation functions such that there is no sampling error at time zero.

In general, to achieve a system-size and time-independent \emph{relative} statistical error using product states, the needed number of samples $S$ could grow with time $t$ for transport-type problems. For example, one could try to sample the infinite temperature average with random product states locally drawn independently from the Haar measure; see Appendix~\ref{sec:haar_product} for an analysis of this ensemble. The sample-to-sample fluctuations are upper bounded by $\braket{Q^\dagger Q}_\infty$, which is provably $O(1)$ for the transport problem. 
However the quantity of interest, $C_{ij}(t)=\braket{Q}_\infty$, is decreasing in time as $t^{-1/z}$. Assuming the $O(1)$ upper bound on the fluctuations is tight, this implies a number of samples growing as $S = O(t^{2/z})$ to obtain a time-independent relative error.

\begin{figure}[t]
\includegraphics[width=\columnwidth]{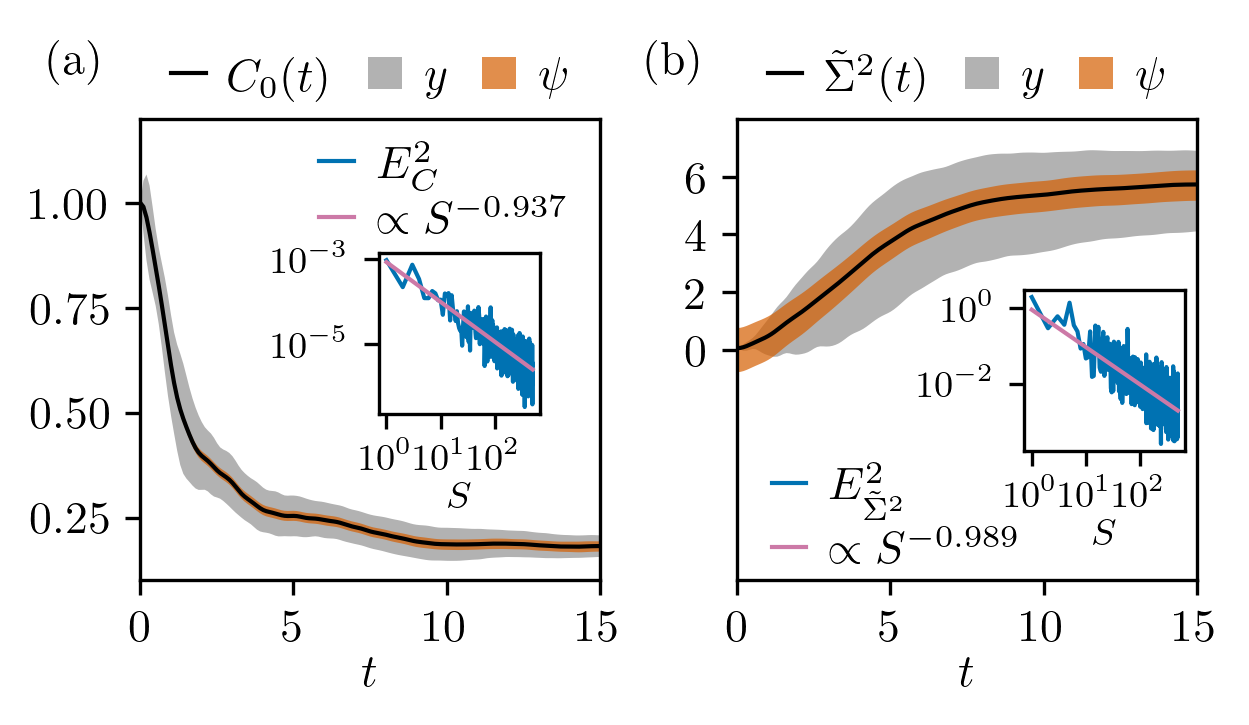}
\centering
\setlength{\abovecaptionskip}{-10pt}
\caption{Exact classical numerical benchmark of sampling $C_0(t)$ and $\tilde{\Sigma}^2(t)$ in the $y$ basis for $L=12$ sites and model parameters $V=1$, $\Omega=2$. Time is measured in units of $1/V$. (a) The exact autocorrelator $C_0(t)$ in black, whereas the gray confidence band is one standard deviation over all $y$-basis product states, named $F_0(L,t)$ in the text. For comparison, the orange band is one standard deviation over $1000$ Haar random states. (b) The same quantities but for $\tilde{\Sigma}^2(t)$. The insets show improvement of accuracy with increasing ensemble size following theoretical predictions, see text for discussion.
}
\label{fig:benc}
\end{figure}

Alternatively, one can harness the phenomenon of quantum typicality to reduce this problem to the calculation of a single expectation value \cite{Goldstein06,Popescu06,gemmer_distribution_2003}. In this method, one draws a random pure state $\ket{\psi}$ from the Haar measure on the entire Hilbert space and finds that the statistical corrections are $ d^{-1/2} \braket{Q^\dagger Q}_\infty$, which is exponentially small in system size since $\braket{Q^\dagger Q}_\infty = O(1)$ for the transport problem. On classical computers, a sample from the Haar ensemble can be generated by fixing some orthonormal basis $\ket{j}$ and directly constructing the entangled state
\begin{equation}
    \left|\psi\right\rangle \propto \sum_{j}c_j\ket{j}
\end{equation}
where $c_j$ are independent and identically distributed (i.i.d.) complex Gaussian random variables with zero mean. This approach of course does not alleviate the exponential cost of storing $\ket{\psi}$ or computing $\braket{\psi|Q|\psi}$, but since it eliminates the need for ensemble averaging, it is still exponentially faster than simulating the entire ensemble. In Fig.~\ref{fig:benc} panel (a) we plot $C_0(t)$ in black, as well as the statistics of $\text{Re} \braket{\psi|h_{L/2}(t) h_{L/2}(0)|\psi}$ over an ensemble of $1000$ Haar random states $\ket{\psi}$, with the orange confidence band corresponding to one standard deviation. We can see that the statistical error is already almost negligible even for $L=12$; a single state suffices to simulate the infinite temperature ensemble.

On quantum computers, the situation is somewhat different. In order to coherently prepare a Haar-random state, at least an exponentially deep local random circuit would be needed~\cite{NielsenChuang}. On the other hand, it may suffice in practice to prepare a sample from an approximate quantum state $k$-design for sufficiently large $k$, which has some statistical properties equivalent to those of a Haar-random state. Actually, these states are proven to require only depth-$O(k^{10}L^2)$ local random quantum circuits to prepare \cite{Brando2016} and are conjectured to require even less depth \cite{hunter-jones_unitary_2019}. This scaling is efficient in principle and the method of using shallow random circuits for spin transport calculations was demonstrated via classical simulations of noisy quantum hardware in Ref.~\cite{Richter21}. This work was followed up by Ref.~\cite{Keenan22} which performed the calculation on IBM hardware and confirmed KPZ scaling for spin transport in the XXZ chain. However, in general on NISQ devices, it may be advantageous to consider methods that skip this state preparation step and save valuable coherence time needed to measure transport behavior.

Having discussed various methods that could be used to sample from the infinite temperature Gibbs ensemble, we note that more direct methods such as operator time-evolving block decimation (TEBD) have also been used to successfully approximate the infinite temperature trace out to long times in restricted Hilbert spaces~\cite{Ljubotina23}.

\subsection{Sampling of Correlators}

Here, we propose to sample from the infinite-temperature ensemble using a \emph{special} set of product states that are tailored to computing energy-density correlation functions in the MFIM. 
In this way, the state preparation step requires only single-qubit gates, making it much more suitable for NISQ devices. 
Let $\ket{y_k}$ be product states where each qubit is in a randomly chosen eigenstate of the Pauli $Y$ operator on the corresponding site.
The subscript $k=1,\dots,S$ indexes $S$ random samples from the set of all $d$ such product states; we denote the corresponding random variable by $\ket{y}$.
We define
\begin{equation}
    C^S_r(t) = \frac{1}{S}\sum_{k=1}^S \text{Re} \braket{y_k|h_{L/2+r}(t)h_{L/2}(0)|y_k}
\end{equation}
to be a sample-averaged energy-density correlator in a random ensemble of $y$-basis states. It follows that statistically,
\begin{equation}\label{eq:fluct}
    C^S_r(t) = C_r(t) + O\bigg(\frac{F_r(L,t)}{S^{1/2}}\bigg)
\end{equation}
where $F_r(L,t)$ is the standard deviation of $\text{Re} \braket{y|h_{L/2+r}(t)h_{L/2}(0)|y}-C_r(t)$ over all $d$ $y$-basis states. Eq.~\eqref{eq:fluct} then follows from the central limit theorem and the fact that $\text{Re} \braket{y|h_{L/2+r}(t)h_{L/2}(0)|y}$ is an unbiased estimator of $C_r(t)$. Of course, the preceding statement holds for any complete basis of states; to advocate for sampling in the $y$-basis versus other bases, we need to say something about the size of $F_r(L,t)$. We demonstrate the following:

\begin{enumerate}
    \item The dynamical relative error $F_0(L,t)/C_0(t)$ is at worst $O(1)$ in $L$ and $t$ and exceptionally small for multiple model parameters.
    
    \item The equivalent dynamical relative error for the $z$ basis can be larger than $100\%$ and appears to increase with $L$ at late times.
    
    \item The initial error $F_r(L,0) = 0$.

    \item The long-time error is exponentially small---i.e., $F_r(L,t\rightarrow \infty) = O(L^\alpha d^{-1/2})$ for some $\alpha>0$---assuming a quantum chaotic system.
    
\end{enumerate}
Points 1 and 2 above are empirical observations from classical numerics, while point 3 is a simple consequence of the relation between the model and the $y$ basis, as discussed below. Point 4 is non-trivial and is demonstrated analytically in Appendix~\ref{sec:justify_y}.

The basic mechanism behind the advantage of sampling in the $y$ basis is that for the Hamiltonian~\eqref{eq:H}, the random product states $\ket{y_k}$ look like 
infinite-temperature Gibbs states as far as one and two point spatial energy-density correlators are concerned:
\begin{align}
    \braket{y|h_i|y} &= \braket{h_i}_{\infty} \quad \forall i,y \\
    \braket{y|h_i h_j|y} &= \braket{h_i h_j}_{\infty} \quad \forall i,j,y.\label{eq:en_en}
\end{align}
An immediate consequence of the second condition is that there is no error at time zero, $F_r(L,0) = 0$. At the same time, we prove in Appendix~\ref{sec:justify_y} that, for a quantum chaotic system,
\begin{equation}
    \lim_{T\rightarrow\infty} \frac{1}{T} \int_0^T dt\  F^2_r(L,t) = O(L^\alpha d^{-1})
\end{equation}
for some $\alpha>0$. To make this statement we assume a mild form of the ``diagonal" eigenstate thermalization hypothesis (ETH)~\cite{Deutsch91,Srednicki94} along with a no-resonance condition on the energy spectrum \cite{srednicki_approach_1999} as well as exponentially small average purity of $y$-basis state diagonal ensembles, all in line with standard assumptions in the study of quantum chaotic systems~\cite{D'Alessio16,riddell_concentration_2022}. Although this bound does not enforce $F_r(L,t)$ to be exponentially small at time scales relevant to diagnosing transport, we find numerically that after an initial transient, the actual error is particularly small at intermediate times (i.e., those from which we extract a dynamical exponent) before it ultimately becomes exponentially small at late times. 

To help demonstrate the sense in which the $y$-basis is empirically useful for this problem, we first perform a classical numerical simulation of the dynamics using a high-order series-expansion approximation to $e^{-iHt}$ which is essentially exact. Fig.~\ref{fig:benc}(a) shows in the gray confidence band the classical simulation of $F_0(L,t)$ for $L=12$. We can see that $F_0(L,t)$ is reasonably small; i.e. a single randomly drawn $y$-basis state more or less captures the physics of the energy transport. To make this more quantitative, in Appendix~\ref{sec:rel_error} we show how $F_0(L,t)/C_0(t)$ scales with $t$ and $L$ for different model parameters and product state bases. We find that the $y$-basis is consistently advantageous. Of course to use this method to identify a dynamical exponent $z$, much more precision is needed than just being within the gray confidence band. Thus, one can average over $S$ random samples---our results suggest that $S$ can be relatively small and, at worst, $O(1)$ in $L$ and $t$.

In principle, Eq.~\eqref{eq:fluct} only holds statistically, and the actual fluctuations of $C^S_r(t)$ around $C_r(t)$ are $O(S^{-1/2})$ in $S$ only when $S$ is large enough that higher moments are suppressed via the central limit theorem. To confirm that the empirical fluctuations are at least in some sense captured by the second moment even for relatively small $S$, we numerically calculate the averaged square deviations of the sample-averaged correlation function $C^S_0(t)$ from the exact $C_0(t)$  as follows:
\begin{align}
    E^2_{C}(S)\equiv\frac{1}{|\mathcal{T}|}\int_{t\in\mathcal{T}}\left|C_{0}^{S}(t)-C_{0}(t)\right|^{2}dt
\end{align}
where the time interval $\mathcal{T}=(0,15)$. In the inset of Fig.~\ref{fig:benc}(a), we can see that indeed $E^2_{C}$ decays with $S$ as $S^{-1}$ up to fluctuations, as indicated by the slope of a best linear fit to $E^2_{C}$ on a log-log scale. The fact that such a clean $S^{-1}$ scaling can be seen can be attributed to the time averaging and to the coarse-graining effected by the least-squares best-fit. The precise form of the scaling aside, the error clearly decreases with $S$.

\subsection{Sampling of Spatial Variance}

The other transport diagnostic that we discussed in Sec.~\ref{Sec:Rydberg} is the spatial variance $\tilde{\Sigma}^2(t)$. Here, we also provide a classical simulation of this quantity along with a benchmark of the $y$-basis method 
as compared to the standard typicality method. Let $C^\psi_r(t) = \text{Re} \braket{\psi|h_{L/2+r}(t)h_{L/2}(0)|\psi}$. In Fig.~\ref{fig:benc}(b), the orange confidence band shows the statistics of
\begin{equation}
    \tilde{\Sigma}^2_\psi(t) = \frac{\sum_r r^2 C_r^\psi(t)}{\sum_r C^\psi_r(t)} - \bigg(\frac{\sum_r r C_r^\psi(t)}{\sum_r C^\psi_r(t)} \bigg)^2
\end{equation}
over $1000$ samples from the Haar ensemble. We can clearly see that the relative error in $\tilde{\Sigma}^2_\psi(t)$ is significantly larger than it was for $C_r^\psi(t)$, which we attribute to the fact that the statistical error in $C_r^\psi(t)$ are amplified by $r^2$. The same feature is observed for the $y$ basis sampling method: let $C^y_r(t) = \text{Re}\braket{y|h_{L/2+r}(t)h_{L/2}(0)|y}$ and define
\begin{equation}
    \tilde{\Sigma}^2_y(t) = \frac{\sum_r r^2 C_r^y(t)}{\sum_r C^y_r(t)} - \bigg(\frac{\sum_r r C_r^y(t)}{\sum_r C^y_r(t)} \bigg)^2.
\end{equation}
The gray confidence band in Fig.~\ref{fig:benc} panel (b) shows the statistics of $\tilde{\Sigma}^2_y(t)$ in the ensemble of all $y$ basis states. This method clearly comes with a large confidence band relative to the value of $\tilde{\Sigma}^2(t)$, but the ratio of the $y$-basis standard deviation to that of the Haar samples is around $3.5$ for both the spatial variance and the autocorrelator; i.e., both methods are more costly for the spatial variance.

To reduce this band, we would like to average over $y$ basis states as we did for $C^S_0(t)$. A distinction 
between $C_r(t)$ and $\tilde{\Sigma}_y^2(t)$, however, is that $\tilde{\Sigma}^2(t)$ is a non-linear function of $\ket{y}$ and thus $\tilde{\Sigma}_y^2(t)$ is a biased estimator of $\tilde{\Sigma}^2(t)$. Interestingly, the nonlinearity is due only to the second term in the spatial variance and not the denominators. This is because the denominators are independent of $\ket{y}$, which follows from Eq.~\eqref{eq:en_en} and energy conservation. While this bias should ultimately be small in a large system since the second term in the spatial variance is present only because of lack of exact reflection symmetry around site $L/2$, we can also mitigate it as follows.

Since the estimation of $\tilde{\Sigma}^{2}(t)$ is an efficient classical post-processing step (scaling as $S\cdot L$), using the data $\{C^{y_k}_r(t)\}_{k=1}^S$ obtained from the QPU, we could first compute all of $\{C_r^S(t)\}_r$ and then use these as data to estimate $\Sigma^2(t)$. Therefore, we define
\begin{equation}\label{eq:approx_unbiased}
    \tilde{\Sigma}^2_S(t) = \frac{\sum_r r^2 C_r^S(t)}{\sum_r C^S_r(t)} - \bigg(\frac{\sum_r r C_r^S(t)}{\sum_r C^S_r(t)} \bigg)^2.
\end{equation}
The denominators are actually independent of $t,y,$ and $S$ in the noiseless case, but we explicitly include these dependencies for future reference; on noisy hardware they will become important. Now, Eq.~\eqref{eq:approx_unbiased} is still a biased estimator of $\tilde{\Sigma}^2(t)$, but the bias decreases as $O(S^{-1/2})$. To numerically demonstrate improvement of the $y$-basis sampling of $\tilde{\Sigma}^2(t)$ with increasing ensemble size in this setting, we also calculate the time averaged square error
\begin{align}
    E^2_{\tilde{\Sigma}^2}(S) \equiv\frac{1}{|\mathcal{T}|}\int_{t\in\mathcal{T}}\left|\tilde{\Sigma}_{S}^{2}(t)-\tilde{\Sigma}^{2}(t)\right|^{2}dt 
\end{align}
and we can see in the inset of Fig.~\ref{fig:benc}(b) that in practice, $E^2_{\tilde{\Sigma}^2}(S)$ behaves approximately as $S^{-1}$.

\section{QPU: Methods and Results}\label{Sec:QC}

\subsection{Measuring Correlation Functions}

To characterize the energy transport, we will need to measure quantities of the form $\braket{\psi|A(t) B(0) |\psi}$ on a quantum computer.
This is a nontrivial task for generic operators $A$ and $B$ because it constitutes an overlap between distinct quantum states $A e^{-iHt}\ket{\psi}$ and $e^{-iHt}B\ket{\psi}$ rather than an expectation value of the form $\braket{\psi|\mathcal O(t)|\psi}$ for some observable $\mathcal O$.
While there are simple, generic quantum computing primitives for computing such overlaps, like the Hadamard~\cite{Ortiz01,Somma02} and swap tests~\cite{Barenco97,Buhrman01}, they require ancilla qubits and/or multiple qubit registers, along with high connectivity controlled-unitary gates that make their naive implementation on quantum hardware challenging.
Some simplifications can be made when the initial state $\ket{\psi}$ is an eigenstate of the observables $A$ and $B$.
For example, when calculating spin transport (see, e.g., Ref.~\cite{joshi_observing_2022}), one can set $A=Z_i$ and $B=Z_j$ and use $z$-basis initial states, in which case the correlation function factorizes as $\braket{\psi|Z_i(t)|\psi}\braket{\psi|Z_j(0)|\psi}$. Again for the case of spin transport, but using a typicality based approach, Refs.~\cite{Richter21,Keenan22} observed that the spin-spin correlator $\braket{Z_i(t)Z_j(0)}$ can be deduced from a one-point function $\braket{Z_i(t)}$ by evolving a pseudo-random state on all sites \emph{except} $j$. However, such simplifications are not directly available for more complicated operators like the energy density operators considered in this work.

To avoid preparing entangled initial states and using costly ancilla-controlled unitaries, we adapt a protocol for replacing Hadamard tests with direct measurements outlined in Ref.~\cite{Mitarai19}, which makes it possible to measure the desired correlation functions when using $y$-basis states. 
We first decompose the energy density correlators in the Pauli basis,
\begin{equation}\label{eq:pauli_decomp}
   h_{L/2+r}(t)h_{L/2}(0) 
   =\sum_{\mu,\nu=1}^{4} \alpha_{\mu,\nu}P_{L/2+r}^{\mu}\left(t\right)P_{L/2}^{\nu}\left(0\right)
\end{equation}
where we have defined 
$P_i^1 = X_i$, $P_i^2=Z_i$, $P_i^3=Z_i Z_{i+1}$, and $P_i^4=Z_{i-1}Z_i$.
The $\alpha_{\mu,\nu}$ are real-valued coefficients that depend principally on $V,\Omega$ but also depend slightly on $r$ since energy densities are defined differently on the boundaries of the chain [see Eq.~\eqref{eq:hi}]. Then, we observe that
\begin{multline}\label{eq:mitarai_fuji}
    \text{Re} \braket{y|P^\mu_{L/2+r}(t) P_{L/2}^\nu(0) |y} \\
    = \frac{1}{2}\braket{+,\nu,y|P^\mu_{L/2+r}(t)|+,\nu,y}  \\ - \frac{1}{2} \braket{-,\nu,y|P^\mu_{L/2+r}(t)|-,\nu,y}
\end{multline}
where $\ket{\pm,\nu,y}=(I \pm P_{L/2}^\nu)\ket{y}/\sqrt{2}$. For a fixed $y$-basis product state and fixed $r$, $\text{Re} \braket{y|h_{L/2+r}(t)h_{L/2}(0)|y}$ is thus computed by preparing the $8$ states $e^{-iHt} \ket{\pm,\nu,y}$ and measuring $P^\mu_{L/2+r}$. Since these operators commute for $\mu \in \{2,3,4\}$ we ultimately need to prepare only $8 \cdot 2= 16$ distinct circuits.

Additionally we observe that, for a Hamiltonian that is reflection invariant around site $L/2$, some of the above circuits are redundant \emph{on the average over $y$-basis states}. Specifically, we have that
\begin{equation}\label{eq:ensemble_symmetry}
    \braket{P^\mu_{L/2+r}(t) P_{L/2}^3(0)}_\infty = \braket{P^\mu_{L/2-r}(t) P_{L/2}^4(0)}_\infty \ \forall \mu
\end{equation}
so that, having estimated the RHS of Eq.~\eqref{eq:ensemble_symmetry} for all $r$, one would also already have the LHS for all $r$, reducing the number of needed circuits from $16$ to $12$ (up to the ensemble average). For an even number of sites, $H$ is not reflection invariant. However, the lack of reflection symmetry around site $L/2$ is due only to a single sub-extensive boundary term; we expect this to become increasingly unimportant for larger and larger systems.

We would like to take advantage of this approximate symmetry to reduce the number of needed circuits, but a potential problem is that we have advocated using only a small ensemble of $S$ $y$-basis states to estimate correlation functions, whereas the symmetry only holds on the average over all $y$-basis states. For an ensemble of $S$ states, the distinction between the LHS and RHS of Eq.~\eqref{eq:ensemble_symmetry} is in principle only $O(S^{-1/2})$. In practice however, for a certain fixed ensemble of $S=12$ $y$-basis states under consideration, we find that by simply assuming
\begin{multline}\label{eq:y_symmetry}
    \braket{y_k|P^\mu_{L/2+r}(t) P_{L/2}^3(0)|y_k} \\
    = \braket{y_k|P^\mu_{L/2-r}(t) P_{L/2}^4(0)|y_k}
\end{multline}
for each $\ket{y_k}$ in that ensemble, we obtain numerical results  very close to those not assuming Eq.~\eqref{eq:y_symmetry}. This is shown and discussed in Appendix~\ref{sec:simplify}.

To further reduce circuit depth and save coherence time for diagnosing transport rather than preparing initial states, we 
make one more simplification.
Note that for $\nu\in\{3,4\}$, the states $\ket{\pm,\nu,y}=(I \pm P_{L/2}^\nu)\ket{y}/\sqrt{2}$ are Bell pairs on sites $L/2\pm 1$ (and product on the other sites). We opt to ignore these superpositions and simply replace these initial states with the local computational (Pauli $Z$) basis states $\{\ket{00},\ket{01},\ket{10},\ket{11}\}$ on sites $L/2-1, L/2$ (for $\nu=3$) and sites $L/2, L/2+1$ (for $\nu=4$), incurring an error which is negligible in practice, as is also shown in Appendix~\ref{sec:simplify}. This replacement comes at the cost of requiring $16$ distinct circuits instead of the $8$ needed when using bell pairs.

\begin{figure}[t]
\includegraphics[width=\columnwidth]{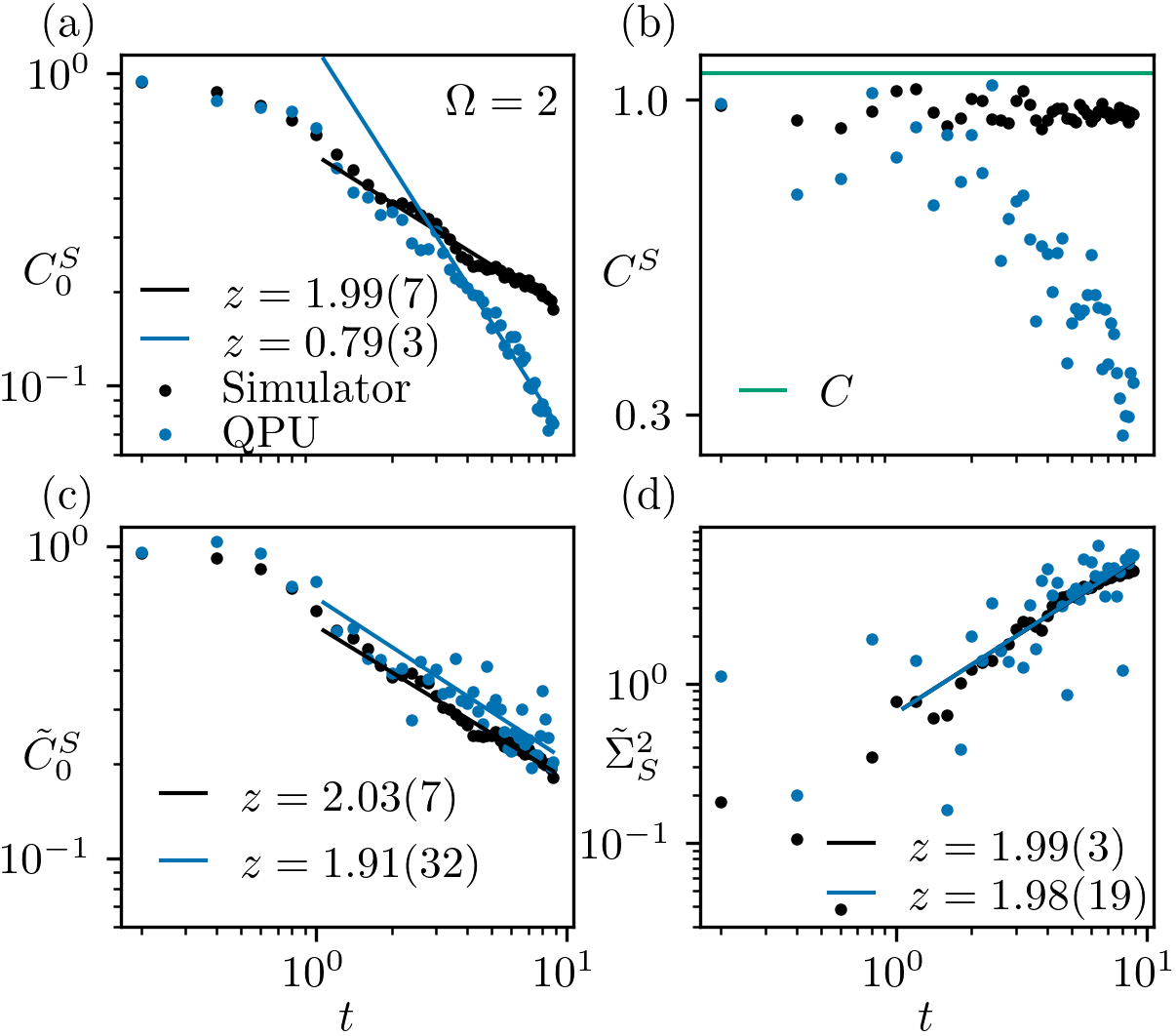}
\centering
\caption{Demonstration of renormalization strategy for simulating energy transport. Black scatter points are obtained from ideal quantum circuit simulations using \texttt{qasm\_simulator} and blue are obtained from quantum hardware experiments on \texttt{ibmq\_montreal}. The black and blue lines are power-law fits to the ideal and QPU results, respectively (using $a t^{-1/z}$ for the energy density autocorrelator and $b t^{2/z}$ for the SV). (a) The $y$-basis-sampled raw energy density autocorrelator $C_0^S(t)$ and (b) their sum $C^S(t)$, shown also with the value $C$ in green for comparison. (c) The renormalized correlator $\tilde{C}_0^S(t)$ and (d) the renormalized spatial variance $\tilde{\Sigma}^2(t)$. For both the simulator and the hardware experiment we set $L=12$, $V=1$, and $\Omega=2$; we fix the Trotter step size $\delta t = 0.1$ and simulate up to $90$ Trotter steps. The data points meaurements at only even Trotter steps. In all simulations we use a fixed sample of $S=12$ $y$-basis initial states which are recorded in Appendix~\ref{sec:simplify}. Quantum expectation values are approximated with $8192$ measurement shots.}
\label{fig:Om_2}
\end{figure}

\subsection{QPU Results}\label{sec:sum_rule}

Here we describe energy transport results obtained on the 27-qubit \texttt{ibmq\_montreal} QPU. We focus on model parameters $L=12$, $V=1$, and $\Omega=2$ in Fig.~\ref{fig:Om_2}. To implement the unitary dynamics $U(t) = e^{-iHt}$ on a digital QPU, we Trotterize the dynamics~\cite{Trotter59,Suzuki76}. One subtlety that arises in the context of energy transport is that energy is no longer strictly conserved with finite Trotter step size. Consider the quantity
\begin{equation}
    C^S(t) = \sum_r C^S_r(t)
\end{equation}
which, without any Trotter error, would be equal to the time independent constant $C$ defined via Eq.~\eqref{eq:const}. This fact follows from Eq.~\eqref{eq:en_en} and energy conservation. The black scatter points in Fig.~\ref{fig:Om_2}(b) are a classical ideal quantum circuit simulation, including shot noise, of the Trotterized dynamics of $C^S(t)$ for $S=12$ $y$-basis states. We can see that $C^S(t)$ initially deviates from $C$ and then slightly fluctuates around a decreased value for later times. The fact that $C^S(t)$ does not significantly decay within the timescale shown implies that any decrease in $C^S_0(t)$ [black scatter points in Fig.~\ref{fig:Om_2}(a)] is due to transport of local energy density and not global non-conservation of energy. In fact, the decay of $C^S(t)$ observed in our QPU results [blue scatter points in Fig.~\ref{fig:Om_2}(b)] serves as a useful diagnostic of the level of noise present on hardware. In Appendix~\ref{sec:noise} we study the behavior of $C^S(t)$ under noisy classical dynamics simulations and find that its decay rate is directly related to the amount of noise in the simulation.

This observation leads to a natural strategy for error mitigation in the context of this particular problem: we renormalize all $L$ correlation functions by their measured sum, defining
\begin{equation}
\label{eq:Ctilde_def}
    \tilde{C}_{r}^{S}(t)\equiv C_{r}^{S}(t)/C^{S}(t).
\end{equation}
In Appendix~\ref{sec:noise} we show data from classical simulations of the Trotter dynamics in the presence of noise and demonstrate the systematic dependence of this strategy's effectiveness on the noise strength.
We find that renormalization provides a substantial improvement up to a time scale (related to the device coherence time) where $C^S(t)$ becomes small. 
The renormalized results begin to become significantly noisy when $C^S(t)$ becomes too small, say less than $0.1$. 
Therefore the size of $C^S(t)$ can be used as a self-consistency check on the quality of the simulation, provided that the decay of the individual $C^S_r(t)$s is sufficiently spatially uniform to justify a uniform global rescaling.
In addition to the renormalization, our QPU runs use standard problem-agnostic error mitigation methods as in our previous work \cite{Chen22}: zero noise extrapolation with scale factors $\{1.0,1.5,2.0\}$, tensored readout error mitigation, dynamical decoupling, and dressing all two-qubit gates with randomly chosen Pauli operators. To significantly save on coherence time, we also implement $R_{ZZ}$ rotations with pulse level control using $R_{ZX}$ rotations native to the device~\cite{Stenger2021}. 

The blue points in Fig.~\ref{fig:Om_2}(a) show unrenormalized QPU results produced using these standard error mitigation techniques.
[The same techniques are also applied to obtain the QPU results in Fig.~\ref{fig:Om_2}(b).]
Clearly, these error mitigation methods alone are insufficient to capture the correct dynamics and predict a different scaling exponent than the ideal circuit results. Fig.~\ref{fig:Om_2}(c) shows, however, that renormalization by $C^S(t)$ in combination with the above methods brings the QPU data points much closer to the ideal simulator, especially at late times.
A power-law fit to the decay of the renormalized energy autocorrelator provides a QPU estimate of $z=2.03$, which is very close to that predicted by the ideal circuit simulator ($z=1.91$).
We can also probe transport via the SV of the energy distribution defined by $\tilde C^S_r(t)$, which we denote by $\tilde{\Sigma}^2_S$.
In Fig.~\ref{fig:Om_2}(d) we see that the QPU and ideal classical simulator both produce an SV growing approximately linearly with $t$. Taken together, these results indicate that the energy transport on timescales probed by these simulations is approximately diffusive.

\begin{figure}[t]
    \centering
    \includegraphics[width=1.0\columnwidth]{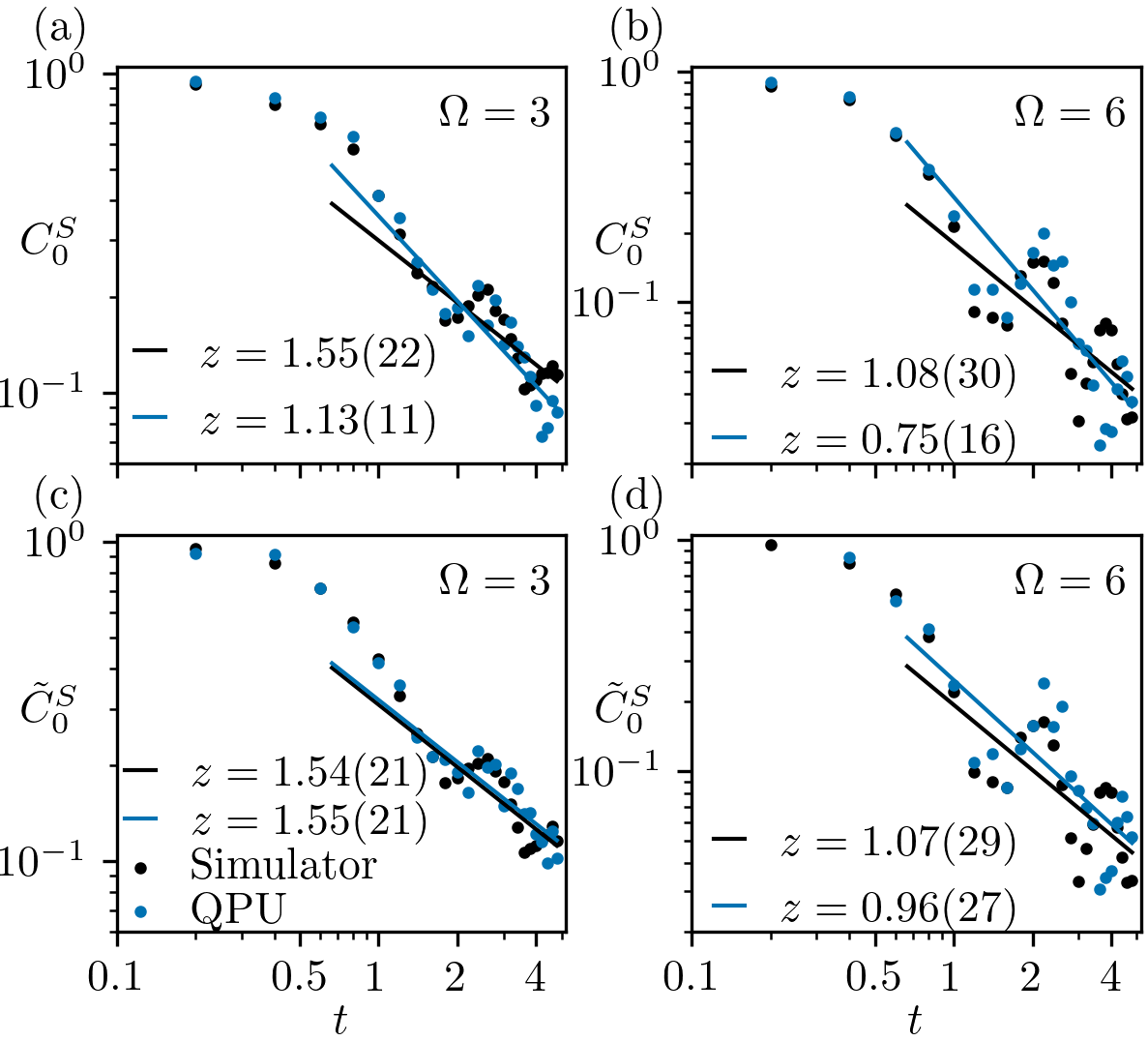}
    \caption{Raw and renormalized energy autocorrelator data from the ideal circuit simulator and the hardware experiment for other $\Omega$. Beyond varying $\Omega$, the model parameters, hyper-parameters and $y$-basis states are identical to Fig.~\ref{fig:Om_2}. The left two panels show the raw (a) and renormalized (c) results at $\Omega=3$ and the right two show $\Omega=6$. Solid lines show fits to $t^{-1/z}$.}
    \label{fig:others}
\end{figure}

As discussed in Sec.~\ref{Sec:Rydberg}, increasing the transverse field strength $\Omega$ brings the Hamiltonian~\eqref{eq:HRydberg} closer to integrability. To investigate the energy transport in this regime and explore how varying model parameters affects the quality of QPU data, we simulate energy transport with the above protocol for $\Omega=3$ and $\Omega=6$. Results are shown in Fig.~\ref{fig:others}. 
Fig.~\ref{fig:others}(a) and (b) show that, even without renormalization, the individual data points obtained from the QPU match the ideal simulator results fairly closely until times $t\sim 3V^{-1}$.
This closer agreement is reasonable given that transport is occurring on shorter timescales for these values of $\Omega$ as compared to the results shown in Fig.~\ref{fig:Om_2}.
Faster transport means that fewer Trotter steps are needed to simulate the correlator's dynamics until it reaches an $O(1/L)$ value, which means lower circuit depth and lower infidelity.
However, the deviations from the simulator results between $t\sim 3$ to $t\sim 5$ are sufficient to skew the QPU estimate of the power-law decay (blue lines) away from the estimates obtained from the ideal simulator (black lines).
In Fig.~\ref{fig:others}(c) and (d), we see that renormalizing the correlator by $C^S(t)$ brings the ideal and QPU results into much closer agreement over the full simulation time window. 
The renormalized QPU data predict transport exponents $z=1.55$ for $\Omega=3$ and $z=0.97$ for $\Omega=6$, consistent with superdiffusive and ballistic behaviors, respectively, that are expected at early times for these parameters.
In Appendix~\ref{sec:OtherParams} we show further QPU data including the SV and spatiotemporal energy correlations.

\section{Generalization to Other Models}\label{sec:generalize}

\subsection{Explicit Examples and a Counterexample}

Before concluding our study of simulating energy transport in the MFIM, we discuss some preliminary results that suggest the method applies more generally. The central idea of our proposal is that, if one can identify a complete basis of product states $\ket{p}$ for which
\begin{equation}\label{eq:general_cond}
    \braket{p|h_i|p} = \braket{h_i}_\infty,\quad \braket{p|h_ih_j|p} = \braket{h_ih_j}_\infty,
\end{equation}
holds for all $\ket{p}$, then this is a particularly good basis in which to sample the infinite temperature dynamical energy-energy correlators. Some other examples besides the MFIM where we can identify the appropriate product-state basis include any odd-body generalization of the $XY$ model, e.g.
\begin{equation}
    h_i = J_{x} X_{i-1} X_i X_{i+1} + J_{y} Y_{i-1} Y_i Y_{i+1},
\end{equation}
where Eqs.~\eqref{eq:general_cond} hold for $z$-basis states. A more physically motivated example is the 1D $\mathbb{Z}_2$ lattice gauge theory coupled to spinless fermions. This theory can be mapped onto a spin chain~\cite{Borla19} with
\begin{equation}
    h_i = J(X_i - Z_{i-1} X_i Z_{i+1}) + h Z_i.
\end{equation}
Here, again the $y$-basis satisfies Eqs.~\eqref{eq:general_cond}. This fact was leveraged in Ref.~\cite{chen_minimally_2024} to study the finite-temperature properties of this model.

A product state basis satisfying \eqref{eq:general_cond} need not exist for a generic Hamiltonian. As an explicit counterexample, consider the isotropic Heisenberg chain for which
\begin{gather}
    h_i = \vec{\sigma}_i \cdot \vec{\sigma}_{i+1},\quad h_i^2 = 3 - 2h_i, \\
    h_i h_{i+1} = \vec{\sigma}_i \cdot \vec{\sigma}_{i+2} + i \ \vec{\sigma}_{i+1} \cdot (\vec{\sigma}_i \times \vec{\sigma}_{i+2}).
\end{gather}
The most general product state basis of three sites is
\begin{equation}
\ket{\vec{a}\pm}\ket{\vec{b}\pm}\ket{\vec{c}\pm}
\end{equation}
where $\vec{a},\vec{b},\vec{c}$ are general unit vectors on the Bloch sphere and $\pm$ refer to the antipodal points along these axes. The desired conditions obtain only if
\begin{equation}
    \vec{a} \cdot \vec{b} = 0,\ \vec{a} \cdot \vec{c} = 0,\  \vec{b}\cdot(\vec{a}\times \vec{c}) = 0,
\end{equation}
which is impossible in three dimensions. This suggests that in general, one will have to relax the assumptions of our sampling approach, for example by allowing for some entanglement in the sampling basis or demanding that Eqs.~\eqref{eq:general_cond} hold only approximately.

\begin{figure*}[t]
    \centering
    \includegraphics[width=0.9\linewidth]{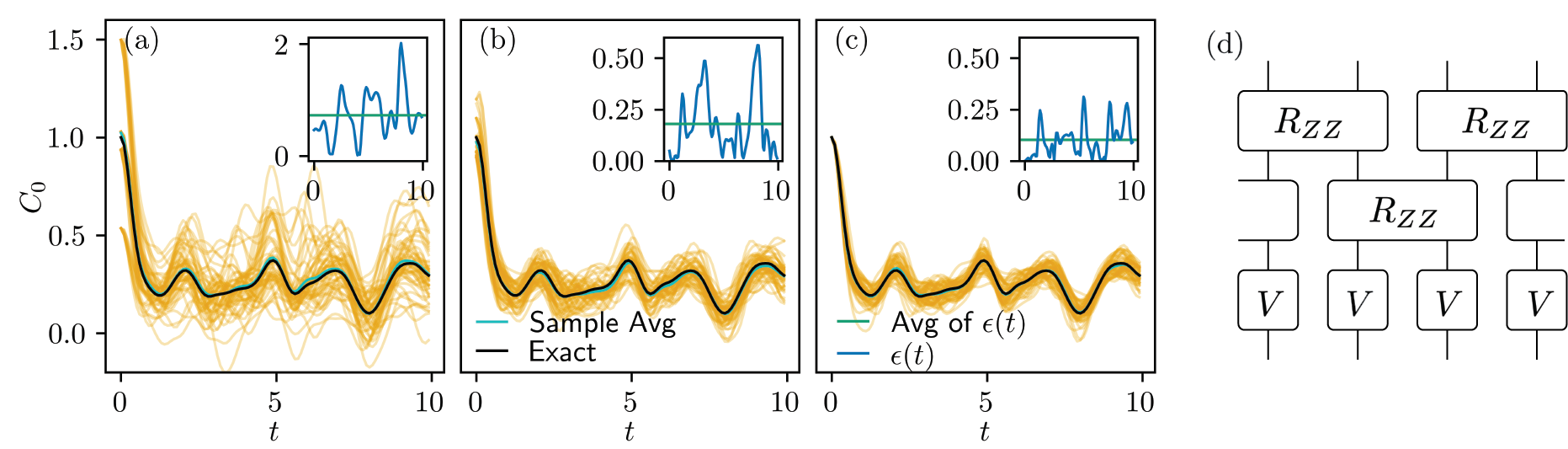}
    \caption{Comparison of sampling methods for an $L=8$-site XXZ spin chain at the Heisenberg point. All plots make use of a variational ansatz circuit of the form depicted in panel (d). (a) Sampled energy density autocorrelator using an optimized product-state basis involving only single-qubit gates $V=R_XR_YR_Z$, corresponding to only the bottom layer of gates depicted in (d). (b) Sampled energy density autocorrelator using a set of random low-entanglement states obtained by feeding random computational basis states into a single instance of the variational ansatz (d) with randomly chosen angles. (c) Sampled energy density autocorrelator using a low-entanglement basis obtained by optimizing the ansatz circuit shown in (d). In all panels, the orange lines represent each of $S=50$ samples obtained by feeding random computational basis states into the respective ansatz circuits, and the black line represents the exact reference dynamics that is well captured by the sample average, shown in cyan. In the insets, the blue lines are the statistical relative error $\epsilon(t)$ and the green horizontal lines are their time-averaged values.}
    \label{fig:gen}
\end{figure*}

\subsection{Variational Approach}

Here, we consider a variational approach to finding a basis which approximately satisfies Eqs.~\eqref{eq:general_cond} while also allowing for some entanglement. First, we fix a shallow entangling circuit ansatz $U(\vec{\theta})$ made of local gates with $O(L)$ variational angles. We also fix a random set of $S$ computational basis states $\ket{k}$. We then minimize the cost function
\begin{align}
    \mathcal{C}_S(\vec{\theta}) = \frac{1}{S}\sum_{k=1}^S \sum_{i} \bigg| \braket{k|U^\dagger(\vec{\theta}) h_i U(\vec{\theta})|k} - \braket{h_i}_\infty \bigg| \\
    + \frac{1}{S}\sum_{k=1}^S \sum_{i,j} \bigg| \braket{k|U^\dagger(\vec{\theta}) h_i h_j U(\vec{\theta})|k} - \braket{h_ih_j}_\infty \bigg| \nonumber
\end{align}
which attempts to simultaneously minimize the amount by which the conditions \eqref{eq:general_cond} are violated. We test the method on the paradigmatic XXZ spin chain with energy density 
\begin{align}
\label{eq:Heisden}
    h_{i}&=\frac{J_{z}}{2}\left(Z_{i}Z_{i+1}+Z_{i-1}Z_{i}\right)\\ \nonumber &\;+\frac{J_{xy}}{2}\left(X_{i}X_{i+1}+Y_{i}Y_{i+1}+X_{i-1}X_{i}+Y_{i-1}Y_{i}\right),
\end{align}
which includes the Heisenberg chain as a special case and therefore does not admit a product state basis satisfying Eqs.~\eqref{eq:general_cond}.
For simplicity, we fix the system size $L=8$ with PBC, set $J_z=J_{xy}=1$ (i.e., tune to the Heisenberg point) and take a fixed set of $S=50$ random initial computational basis states. We use the shallow circuit ansatz shown in Fig.~\ref{fig:gen}(d) which contains $R_{ZZ}$ rotations and single-qubit rotations $V = R_XR_YR_Z$ for a total of $4L$ variational angles. The SciPy COBYLA optimizer was used to carry out the optimization. Once the optimal $U$ is found, we can sample the correlator $C_0(t)$ via
\begin{equation}
    C_0^k(t)=\text{Re} \braket{k|U h_{L/2}(t) h_{L/2}(0)U|k}.
\end{equation}
Fig.~\ref{fig:gen}(c) shows the result of this method, and panels (a) and (b) are different methods for comparison. Panel (a) shows the same optimization, but using only the single qubit layer of $V$'s and no entangling gates. Panel (b) shows the result of using a randomly chosen set of fixed angles within the same shallow ansatz circuit of panel (d). The orange curves in each panel indicate individual samples $C_0^k(t)$.

Qualitatively, the optimized entangled basis performs the best among all three variants---the samples are better concentrated around the exact dynamics. Allowing some entanglement though is crucial; the optimized product basis in panel (a) would require $100$ times the number of samples to achieve the same accuracy as even the naive shallow random circuit in panel (b) which involves no optimization. To be more quantitative, we also plot in the inset of each panel the corresponding dynamical relative error $\epsilon(t) = F_0(t)/C^S_0(t)$, where
\begin{gather}
    C^S_0(t) = \frac{1}{S} \sum_{k=1}^S C^k_0(t) \\
    F^2_0(t) = \frac{1}{S} \sum_{k=1}^S |C_0^k(t) - C^S_0(t)|^2.
\end{gather}
We see that the optimized entangled sampling basis performs best, with the smallest overall time-averaged relative error (green lines). We find that the time-averaged relative error of the optimized entangled ansatz state in panel (c) is about 0.56 times that of the unoptimized ansatz state with randomly chosen angles in panel (b).
This implies that the optimized basis would require about four times fewer samples than the random circuit for the same overall accuracy.
This preliminary analysis suggests that the principles underlying our sampling approach for the MFIM can be generalized to study energy transport for arbitrary Hamiltonians.

\section{Conclusion and Outlook}\label{Sec:conclude}

In this work, we have proposed a sampling approach for approximating energy-density correlation functions in the MFIM. The method relies on the fact that every $y$-basis product state looks like a spatially uniform infinite temperature Gibbs state as measured by quadratic functions of energy-density operators; we have shown analytically that this leads to an exponentially small statistical sampling error at long times in the strongly chaotic parameter regime of the MIFM. We have also shown numerically that the exponentially small long-time value is accompanied by an empirically small statistical error at intermediate times and for two sets of model parameters (see Appendix~\ref{sec:justify_y}). 

Using this approach, we simulated energy transport in the $L=12$ MFIM on the 27 qubit QPU \texttt{ibmq\_montreal}. We found that the most effective error mitigation strategy is the problem-aware approach of renormalizing correlation functions by their sum, which corrects for global energy non-conservation due to decoherence. From QPU simulations of the energy density autocorrelation function in the middle of the chain, we obtained intermediate-time dynamical exponents very close to those predicted by an ideal quantum circuit simulator. We extracted from the QPU data the values $z=2.03$ for transverse field $\Omega=2$ corresponding to the far from integrable regime, as well as $z=1.55$ for $\Omega=3$ and $z=0.97$ for $\Omega=6$. The latter two cases demonstrate that, qualitatively and at relatively early times, the model displays energy transport increasingly close to ballistic as $\Omega$ increases.

Digital quantum simulation of energy transport provides a potential route to demonstrating practical quantum advantage: one could try to generalize the $y$-basis method to two dimensions and apply it to a larger and newer (with longer coherence times) quantum devices (e.g.~the 127 qubit \texttt{ibm\_kyiv} or 433 qubit \texttt{ibm\_seattle}), and estimate energy correlators out to times longer than are accessible with TN or other approximate methods~\cite{Tindall23,Begusic23}. The renormalization strategy would play an essential role in reaching sufficiently late times. At the same time, the fact that the ideal energy correlators obey a spatial sum rule means the constancy of their sum could be used as a metric to asses the quality of a given quantum hardware platform.

To assess the viability of generalizing our sampling approach to other models, we have performed a preliminary numerical calculation for the isotropic XXZ spin chain by variationally identifying an ensemble of weakly entangled states which look only approximately like infinite-temperature states for one and two point functions of the energy density. We found significantly reduced sampling costs compared to sampling from product states. A worthwhile topic for future work is to formulate the optimization in a more rigorous state-independent way, i.e. in terms of a variational unitary $U(\vec{\theta})$ and the energy density operators alone. The analytical analysis in Appendix~\ref{sec:justify_y} could also be extended to a finitely-entangled sampling basis generated by a depth-$O(1)$ circuit.

Energy transport at \emph{finite} temperature is also of theoretical and experimental interest. The $y$-basis method may also be helpful in that context, and future work could analyze how the sampling complexity depends on $L,t$, and inverse temperature $\beta$ in a manner similar to Ref.~\cite{Saroni23}. One can also ask if the $y$-basis is useful for sampling equilibrium finite-temperature expectation values of local observables via imaginary-time evolution of $y$-basis states~\cite{qite_chan20,VQITE,AVQITE,Sun2021QuantumCO,AVQMETTS}.

\section*{Data Availability}

An example code demonstrating the method adopted in this paper in a noisy simulator is publicly available at \url{https://gitlab.com/yxphysics/diffusion}.

\acknowledgments

The authors acknowledge valuable discussions with Joe Bhaseen, Anatoly Dymarsky, Joseph Eix, Antonio Anna Mele, Antonio Mezzacapo, Mario Motta, Sona Najafi, Jonathon Riddell, Derek Wang, Yi-Zhuang You, and Marko Znidari\v{c}. This material is based upon work supported by the National Science Foundation under Grant No.~DMR-2038010 (K.P., P.P.O., T.I.). Calculations on quantum hardware were supported by the U.S. Department of Energy, Office of Science, National Quantum Information Science Research Centers, Co-design Center for Quantum Advantage (C2QA) under contract number DE-SC0012704 (Y.-X.~Y.). Part of the research (I.-C.~C.) was supported by the U.S. Department of Energy (DOE), Office of Science, Basic Energy Sciences, Materials Science and Engineering Division. This part of the research was performed at the Ames National Laboratory, which is operated for the U.S. DOE by Iowa State University under Contract No. DE-AC02-07CH11358.

\begin{appendix}

\section{Theoretical justification for Y-basis product state sampling}\label{sec:justify_y}

In this appendix we argue that, in a system which obeys a no-resonance condition on the energy spectrum as well as the ETH, the infinite-temperature correlation functions $C_r(t)$ are particularly well-sampled by the special product states $\ket{y}$. More specifically, we show that in an ETH system, the statistical fluctuations of $\braket{y|h_i(t)h_j(0)|y}$ around the exact value are \emph{on the average over a long time} exponentially small in the system size $L$. While this fact certainly does not imply that at all times $t$ the statistical fluctuations are small, we give numerical evidence that after a short initial transient, the effect of the exponentially small long-time value is already felt in the sense that the statistical $y$-basis fluctuations are already particularly small. We further this claim by showing, numerically, a significant difference in the dynamical relative error for $y$-basis sampling as compared to another basis, such as the $z$-basis, which has no special properties with respect to the models under consideration.

In this appendix we will still focus on the MFIM, but without necessarily imposing the constraint that the $ZZ$ coupling and $Z$ field are related as is done implicitly in Eq.~\eqref{eq:HRydberg}. We now take
\begin{equation}\label{eq:en_dense_appendix}
    h_j = N^{-1} [h_x^j X_{j}+h_z Z_{j}+\frac{V}{2}(Z_{j-1}Z_{j}+Z_{j}Z_{j+1})]
\end{equation} in the bulk  $(2\leq j\leq L-1)$ and
\begin{equation}
\label{eq:h_edge}
h_{j} = \frac{1}{N} \begin{cases}
h_x^j X_{1}+\frac{h_z}{2} Z_{1}+\frac{V}{2}Z_{1}Z_{2}& j=1\\
h_x^j X_{L}+\frac{h_z}{2} Z_{L}+\frac{V}{2}Z_{L-1}Z_{L} & j=L\\
\end{cases}
\end{equation}
at the boundaries. We set $N = (V^2/2 + h_x^2+h_z^2)^{1/2}$. For the purposes of this appendix, we will consider two sets of model parameters:
\small
\begin{align*}
    (V,h_x^j,h_z) =(1,2,2) & \indent  \text{(``QPU Experiment")} \\
    (V,h_x^j,h_z) = (1,-1.05+r_j,0.5) & \indent \text{(``Strongly Chaotic")},
\end{align*}
\normalsize
where $r_j \in [-0.01,0.01]$ is a small uniformly distributed random perturbation to remove any exact degeneracy for technical reasons; this way $E_n=E_m$ if and only if $n=m$. Beyond this, we ignore the $r_j$ in our subsequent analysis, e.g.  we take $\braket{h^2_j}_\infty = 1$.

\subsection{Long time averaged fluctuations}

Consider the quantity
\begin{equation}\label{deviations}
    \Delta^{ij}_y(t) = \braket{y|h_i(t)h_j(0)|y} - \langle h_i(t)h_j(0) \rangle_{\infty}.
\end{equation}
This has the property that
\begin{equation}
    \mathbb{E}_y\big[\Delta^{ij}_y(t)\big] = 0 \quad \forall i,j,t.
\end{equation}
where $\mathbb{E}_y = d^{-1}\sum_y$ and $d=2^L$ is the Hilbert space dimension. In other words, random product states are unbiased estimators of a trace. We would like to study the statistical fluctuations of this quantity across randomly drawn $y$-basis product states $\ket{y}$. If the fluctuations are small, then $\langle h_i(t)h_j(0) \rangle_{\infty}$ could be accurately estimated with a small ensemble of random product states. Ideally, we would like to show that 
\begin{equation}
    \mathbb{E}_y\big[|\Delta^{ij}_y(t)|^2\big]
\end{equation}
is small for all times $t$. Note that in the main text we discussed the quantity
\begin{equation}
\label{eq:F2r_initial_bound}
    F^2_{r}(L,t) = \mathbb{E}_y \big[\text{Re} \Delta^{ij}_y(t)\big]^2\leq \mathbb{E}_y|\Delta^{ij}_y(t)|^2
\end{equation}
(here $i=L/2$ and $j=L/2+r$), but in this appendix we focus on bounding the sum of the fluctuations of the real and imaginary parts for analytical simplicity. Despite this simplification, we will see that this simpler bound is still ``qualitatively tight" in the sense that it scales with $L$ as do the fluctuations of the real part of the error. 

Faced with the lack of a closed form expression for $h_i(t)$, as in the original problem of estimating the dynamical exponent, we opt to make some statements instead about the long-time average
\begin{equation}
    \overline{\mathbb{E}_y\big[|\Delta^{ij}_y(t)|^2\big]} = \lim_{T\rightarrow\infty} \frac{1}{T}\int_0^T dt \ \mathbb{E}_y\big[|\Delta^{ij}_y(t)|^2\big].
\end{equation}
To do this, we assume the standard no-resonance condition \cite{srednicki_approach_1999}
\begin{equation}\label{eq:fluct_time_avg}
    \lim_{T\rightarrow\infty} \frac{1}{T}\int_0^T dt\  e^{i(E_n-E_m-E_k+E_l)t} = \delta_{nm}\delta_{kl} + \delta_{nk}\delta_{ml}.
\end{equation}
Recent work has actually revealed that this standard assumption need not hold exactly, independently of chaos via standard indicators \cite{riddell_no-resonance_2023}. However, the corrections to this formula are expected to be subdominant, making it sufficient for our purposes. Inserting resolutions of identity in the eigenbasis of $H$ and making use of the no-resonance condition, we get
\begin{multline}\label{eq:fluct_time_avg_en_basis}
    \overline{\mathbb{E}_y\big[|\Delta^{ij}_y(t)|^2\big]} \\
    = \mathbb{E}_y \bigg| \sum_{n} \braket{n|h_i|n} \bigg(\braket{n|h_j|y}\braket{y|n} - \frac{\braket{n|h_j|n}}{d}  \bigg ) \bigg|^2 \\
    + \mathbb{E}_y \sum_{nm} |\braket{n|h_i|m}|^2 \bigg|\braket{m|h_j|y}\braket{y|n} - \frac{\braket{m|h_j|n}}{d}  \bigg|^2.
\end{multline}
Expanding the square in the second term of Eq.~\eqref{eq:fluct_time_avg_en_basis} and using that $\ket{y}$ is a complete basis allows this term to be written as
\begin{multline}\label{eq:fluct_time_avg_en_basis_sec_term}
    \mathbb{E}_y \sum_{nm} |\braket{n|h_i|m}|^2 |\braket{m|h_j|y}|^2 |\braket{y|n}|^2 \\
    - \frac{1}{d^2} \sum_{nm} |\braket{n|h_i|m}|^2 |\braket{n|h_j|m}|^2.
\end{multline}
Looking towards an upper bound, we can drop the negative term. However this is still an asymptotically tight upper bound because $\braket{h^2_j}_\infty=1$ and $||h_j||^2=O(1)$ where $||\cdot||$ is the operator norm (maximum singular value) and the second term is $O(d^{-1})$. The first term in Eq.~\eqref{eq:fluct_time_avg_en_basis_sec_term} is further upper bounded as
\begin{equation}\label{eq:with_iprs}
    \frac{1}{d}\sum_{nm} |\braket{n|h_i|m}|^2 \bigg(\mathcal{I}_y(\ket{n}) \mathcal{I}_y(h_j\ket{m})\bigg)^{1/2}
\end{equation}
where we have defined
\begin{equation}
    \mathcal{I}_y(\ket{\psi}) \equiv \sum_y |\braket{y|\psi}|^4
\end{equation}
to be the inverse participation ratio (IPR) of the state $\ket{\psi}$ in the $y$ basis. 

\begin{figure*}[t]
\centering
\setlength{\abovecaptionskip}{-5pt}
\includegraphics[width=2.05\columnwidth]{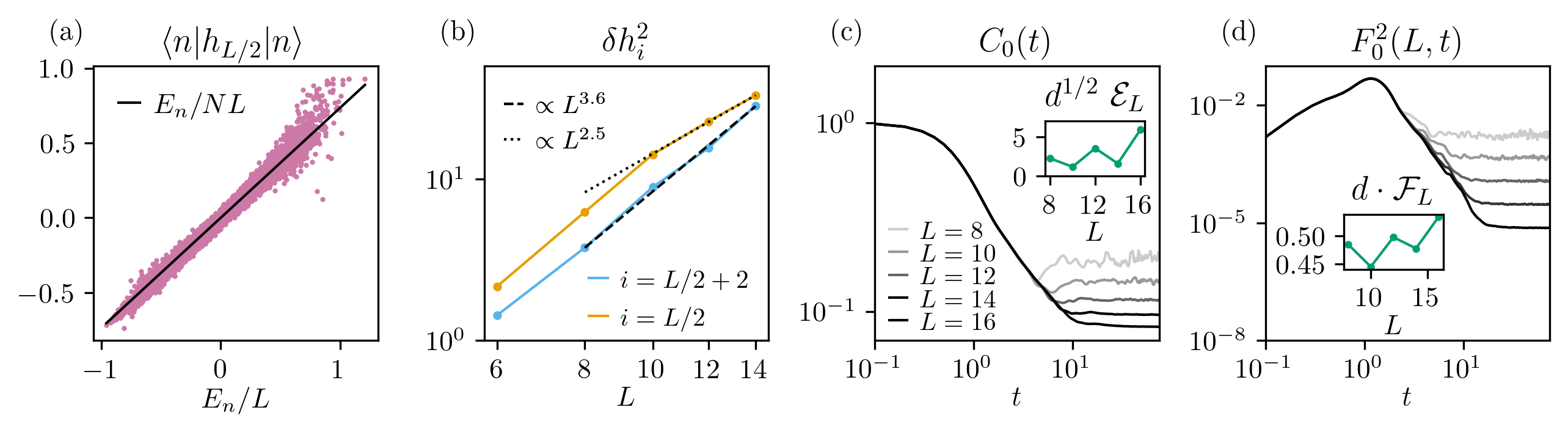}
\caption{Eigenstate thermalization and its consequences for long-time values of dynamical quantities of interest. (a) The diagonal matrix elements of $h_{L/2}$ concentrating around their thermal value for $L=14$. (b) Scaling with $L$ of $\delta h_i^2$ for $i=L/2$ and $L/2+2$. Data are consistent with the theoretical expectation that $\delta h_i^2$ is only polynomially large in an ETH-obeying system. (c) Exact energy-density autocorrelator $C_0(t)$ for system sizes up to $L=16$. The inset shows evidence that $C_0(t\rightarrow \infty)$ deviates from the semi-classical value by an exponentially small amount, see Eq.~\eqref{eq:def_al}. (d) Statistical error in using $y$ basis states to sample the correlator $C_0(t)$. The inset shows that $F_0(L,t\rightarrow\infty)$ is also exponentially small up to a prefactor, see Eq.~\eqref{eq:def_bl}. } 
\label{fig:eth_stuff}
\end{figure*}

To simplify Eq.~\eqref{eq:with_iprs}, we observe that \begin{equation}\label{eq:ipr_bound}
    \mathcal{I}_{y}(h_j\ket{\psi}) \leq 64 \ \mathcal{I}_y({\ket{\psi}})
\end{equation}
for any state $\ket{\psi}$ and in particular, for $\ket{\psi} = \ket{n}$. In other words, if $\ket{\psi}$ is delocalized in a product state basis like $\ket{y}$, a local operator $h_j$ cannot parametrically localize $\ket{\psi}$ in that basis. Eq.~\eqref{eq:ipr_bound} follows precisely from the locality of $h_j$ and the fact that $\ket{y}$ are product states. We prove Eq.~\eqref{eq:ipr_bound} as follows. Inserting a complete $y$-basis into the IPR yields
\begin{equation}
    \mathcal{I}_{y}(h_j\ket{\psi}) = \sum_{y_A y_B} \bigg| \sum_{y'_A y'_B} \braket{y_A y_B|h_j|y'_A y'_B} \braket{y'_A y'_B|\psi}\bigg|^4
\end{equation}
where $\ket{y_A y_B} = \ket{y_A}\ket{y_B}$ with $A$ being the subset of the chain consisting of sites $j-1,j,j+1$ and $B$ the rest, so that $y_A(y_B)$ labels the possible $y$-basis product states within region $A(B)$. It is then clear that
\begin{equation}
    \mathcal{I}_{y}(h_j\ket{\psi}) = \sum_{y_A y_B} \bigg| \sum_{y'_A} \braket{y_A|h_j|y'_A} \braket{y'_A y_B|\psi}\bigg|^4.
\end{equation}
Applying Cauchy-Schwartz inequality to the sum over $y'_A$ then employing the fact that $\braket{y_A|h^2_j|y_A} = 1$ for any $y_A$, we find the bound
\begin{equation}
     \mathcal{I}_{y}(h_j\ket{\psi}) \leq 8 \sum_{y_B y_A y'_A} |\braket{y_A y_B|\psi}|^2 |\braket{y'_A y_B|\psi}|^2.
\end{equation}
Using inequality of geometric and arithmetic means (GM-AM) gives
\begin{align}
     \mathcal{I}_{y}(h_j\ket{\psi}) &\leq 8 \sum_{y_B y_A y'_A} \frac{|\braket{y_A y_B|\psi}|^4 + |\braket{y'_A y_B|\psi}|^4}{2} \\
     &= 8 \sum_{y_B y_A y'_A} |\braket{y_A y_B|\psi}|^4
\end{align}
from which Eq.~\eqref{eq:ipr_bound} follows. Using this result, we may upper bound Eq.~\eqref{eq:with_iprs} by
\begin{align}
     \frac{8}{d}\sum_{nm} |\braket{n|h_i|m}|^2 \big[\mathcal{I}_y(\ket{n})\mathcal{I}_y(\ket{m})\big]^{1/2} \\
     \leq 8||h_i||_{\infty} \braket{\mathcal{I}_y}_\infty 
\end{align}
where in the second line we used again the GM-AM inequality and defined
\begin{equation}
    \braket{\mathcal{I}_y}_\infty = \frac{1}{d}\sum_n \mathcal{I}_y(\ket{n})
\end{equation}
to be the spectrally averaged $y$-basis IPR. Eq.~\ref{eq:with_iprs} is thus bounded by the average IPR of eigenstates in the $y$-basis up to an $O(1)$ factor. The numerical factor of $8||h_i||_{\infty}$ likely leads to a loose upper bound as compared to numerics at small system sizes, but we nonetheless rely on this result because it rigorously implies a suppression of the second term contributing to Eq.~\eqref{eq:fluct_time_avg_en_basis} by the average inverse participation ratio.

Now, let us turn to the first term contributing to Eq.~\eqref{eq:fluct_time_avg_en_basis}. We express $h_i$ as two terms:
\begin{equation}\label{eq:eth_form}
    h_i = \frac{1}{N}\frac{H}{L} + \tilde{h}_i
\end{equation}
where $N$ is the normalization factor introduced in Eq.~\eqref{eq:hi} and $\tilde{h}_i$ is a Hermitian operator. We make this ansatz for $h_i$ because we expect that most matrix elements of $\tilde{h}_i$ will be small in a system obeying the eigenstate thermalization hypothesis (ETH). We discuss the ETH in more detail in the next subsection, but for now let us treat Eq.~\eqref{eq:eth_form} as a definition of the operator $\tilde{h}_i$ whether or not ETH holds for the system under consideration. Plugging this into the first term in Eq.~\eqref{eq:fluct_time_avg_en_basis}, we find
\begin{equation}
    \mathbb{E}_y \bigg| \sum_{n} \braket{n|\tilde{h}_i|n} \bigg(\braket{n|h_j|y}\braket{y|n} - \frac{\braket{n|h_j|n}}{d}  \bigg ) \bigg|^2
\end{equation}
where the term proportional to $E_n$ has dropped out because Eq.~\eqref{eq:en_en} holds for each $y$. Expanding out the square, we find this expression is upper bounded by
\begin{multline}\label{eq:diag_term_bound}
    \mathbb{E}_y \sum_{n} |\braket{n|\tilde{h}_i|n}|^2 |\braket{y|n}|^2 \braket{y|h^2_j|y} \\
    -  \frac{1}{d^2} \sum_n |\braket{n|\tilde{h}_i|n}|^2 \sum_n |\braket{n|h_j|n}|^2.
\end{multline}
Now letting
\begin{equation}\label{eq:deltah_def}
    \delta h_i = \bigg( \sum_n |\braket{n|\tilde{h}_i|n}|^2\bigg)^{1/2},
\end{equation}
we find that Eq.~\eqref{eq:diag_term_bound} is bounded by
\begin{equation}
    \frac{\delta h_i^2}{d} \bigg\{ 1 - \bigg[ \bigg( \frac{\braket{H^2}_\infty}{N^2L^2} \bigg)^{1/2} + \bigg(\frac{\delta h_j^2}{d} \bigg)^{1/2} \bigg]^2 \bigg\}.
\end{equation}
We thus conclude this section with the upper bound
\begin{multline}
\label{eq:intermediate_bound}
    \overline{\mathbb{E}_y\big[|\Delta^{ij}_y(t)|^2\big]} \leq O\big(\braket{\mathcal{I}_y}_\infty\big) \\
    + \frac{\delta h_i^2}{d} \bigg\{ 1 - \bigg[ \bigg( \frac{\braket{H^2}_\infty}{N^2L^2} \bigg)^{1/2} + \bigg(\frac{\delta h_j^2}{d} \bigg)^{1/2} \bigg]^2 \bigg\}.
\end{multline}
In the next subsections we will see that, up to a multiplicative polynomial prefactor, this entire expression is exponentially small in system size.

\subsection{Assuming the ETH}

Srednicki \cite{srednicki_approach_1999} proposed that in a non-integrable system and for a local operator such as $h_i$,
\begin{equation}\label{eq:srednicki}
    \braket{n|h_i|m} = h(\bar{E}/L)\delta_{nm} + \mathcal{D}^{-1/2}(\bar{E}) f(\bar{E},\omega) R_{nm}
\end{equation}
where $\bar{E} = (E_n+E_m)/2$, $\omega = E_n-E_m$, $f$ and $h$ are smooth functions, $\mathcal{D}(\bar{E})$ the smeared density of states at energy $\bar{E}$, and $R_{nm} = O(1)$. Since $h_i$ is an energy density operator, we identify $h(\bar{E}/L)$ with the semi-classical value of $\bar{E}/NL$. In a quantum chaotic system, we expect $f(eL,\omega)$ for fixed arguments $e,\omega$ to scale with $L$ only polynomially, since any exponential dependence has been stripped off via the factor $\mathcal{D}(E)^{-1/2}$.

If we assume the ETH as in Eq.~\eqref{eq:srednicki} and treat the spectum as continuous, we observe that (recall Eq.~\eqref{eq:deltah_def})
\begin{equation}\label{eq:integral}
    \delta h_i^2 = \int dE |f(E,0)|^2,
\end{equation}
so that $\delta h_i^2$ should scale polynomially in $L$. In Fig.~\ref{fig:eth_stuff}(a) we can see that almost all diagonal matrix elements of $h_{L/2}$ are concentrated around their semi-classical value. Fig.~\ref{fig:eth_stuff}(b) shows numerical evidence that $\delta h_i^2$ is bounded by a power law. The $L$ scaling of the ETH function $f(E,\omega)$ for $\omega\rightarrow 0$ has been discussed by assuming hydrodynamics at late times in various types of systems \cite{D'Alessio16,dymarsky_new_2019,capizzi_hydrodynamics_2024}. Such arguments can, however, break down for very small omega, for example $\omega \lesssim \tau^{-1}$ where $\tau = O(L^z)$ is the thermalization time. In any case, the zero frequency behavior at finite $L$ should diverge at most as a power of $L$. Integrating over $E$ may increase this power, but still maintain a polynomial scaling. We therefore assume that there exists some $\alpha>0$ such that
\begin{equation}\label{eq:delta_h_assume}
    \delta h_i^2 = O(L^\alpha).
\end{equation}

We highlight two consequences of Eq.~\eqref{eq:delta_h_assume} here. Firstly, we note that the long-time value of correlation functions for $|r|<L/2$ are given by $(NL)^{-1}\braket{H^2}_\infty$ up to exponentially small corrections. This is because the deviations from this long-time value are bounded as
\begin{multline}
    \bigg|\overline {C_r(t)} - \frac{\braket{H^2}_\infty}{N^2L^2}\bigg|\\ \leq \bigg(\frac{\braket{H^2}_\infty}{N^2L^2}\bigg)^{1/2} (\delta h_i + \delta h_j) d^{-1/2} + \delta h_i \delta h_j d^{-1}
\end{multline}
so that Eq.~\eqref{eq:delta_h_assume} implies
\begin{align}
    \overline{C_r(t)} &= \frac{\braket{H^2}_\infty}{N^2L^2} + O\bigg(\frac{L^{(\alpha-1)/2}}{d^{1/2}}\bigg)\label{eq:eth_pred}  \\ &=  \frac{1}{L} + O(L^{-2}) + O\bigg(\frac{L^{(\alpha-1)/2}}{d^{1/2}}\bigg).\label{eq:semi-classical}
\end{align}
This latter equation is consistent with the semi-classical picture that energy density initially localized on site $i$ has spread roughly uniformly across the system. The second important consequence is that [combining Eqs.~\eqref{eq:F2r_initial_bound}, \eqref{eq:intermediate_bound}, and \eqref{eq:delta_h_assume}]
\begin{equation}\label{eq:final_bound}
    \overline{F^2_r(L,t)} = O\big(\braket{\mathcal{I}_y}_\infty\big) + O\bigg(\frac{L^\alpha}{d}\bigg)
\end{equation}
suggesting that so long as $\braket{\mathcal{I}_y}_\infty$ is exponentially small, so too is the long-time statistical error in using a $y$-basis state to sample the infinite temperature energy-density correlators.

In Fig.~\ref{fig:eth_stuff}(c) and (d) we test these predictions numerically. Clearly panel (c) is consistent with Eq.~\eqref{eq:semi-classical}, i.e. that $C_0(t)$ decreases with $L$ at long times. More quantitatively, let $\mathcal{T} = (12,75)$ be the time averaging window within which we assume that the long-time value has already been reached for $L\leq 16$. The panel (c) inset shows that the error between the theoretical long time value and the actual late time average
\begin{equation}\label{eq:def_al}
    \mathcal{E}_L = \frac{1}{|\mathcal{T}|}\int_{t\in\mathcal{T}} dt \ C_0(t) -\frac{\braket{H^2}_\infty}{N^2L^2} 
\end{equation}
is, up to some weakly increasing prefactor [as predicted by Eq.~\eqref{eq:eth_pred}], proportional to $d^{-1/2}$. As for the $y$-basis error, Fig.~\ref{fig:eth_stuff}(d) suggests there is indeed an exponentially decreasing long time value, and the inset demonstrates that the actual late time average 
\begin{equation}\label{eq:def_bl}
    \mathcal{F}_L = \frac{1}{|\mathcal{T}|}\int_{t\in\mathcal{T}} dt \ F_0(L,t)^2
\end{equation}
is basically proportional to $d^{-1}$ up to a weakly growing prefactor, as is theoretically allowed for in Eq.~\eqref{eq:final_bound}. These numerical results can also be seen as evidence that $\braket{\mathcal{I}_y}_\infty$ is $O(d^{-1})$, up to a weakly growing prefactor. In the next section, we give more direct numerical evidence of this.

\subsection{Inverse participation ratio}\label{sec:IPR}

In the previous subsection, we argued that up to exponentially small corrections, the infinite-time averaged fluctuations are bounded by something of order the average IPR of $y$-basis states in the eigenbasis of $H$, for the MFIM. On a heuristic level, we expect this spectral average IPR to be exponentially small in a chaotic system since
\begin{equation}
    \braket{\mathcal{I}_y}_\infty = \frac{1}{d}\sum_y \sum_n |\braket{y|n}|^4 = \mathbb{E}_y \text{tr}(\omega^2_y)
\end{equation}
where
\begin{equation}
    \omega_y = \sum_n |\braket{y|n}|^2 \ket{n} \bra{n}
\end{equation}
is the diagonal ensemble corresponding to a $y$-basis product state $\ket{y}$. The purity of the diagonal ensemble of a product state in a chaotic system is generically expected to be exponentially small in $L$ \cite{riddell_concentration_2022} and in this case since it is averaged over all $y$-basis states we expect it to be particularly well behaved. However, as a matter of possibly independent interest, we further study the relationship between eigenstates of the MFIM and $y$-basis product states.

\begin{figure}
\includegraphics[width=\columnwidth]{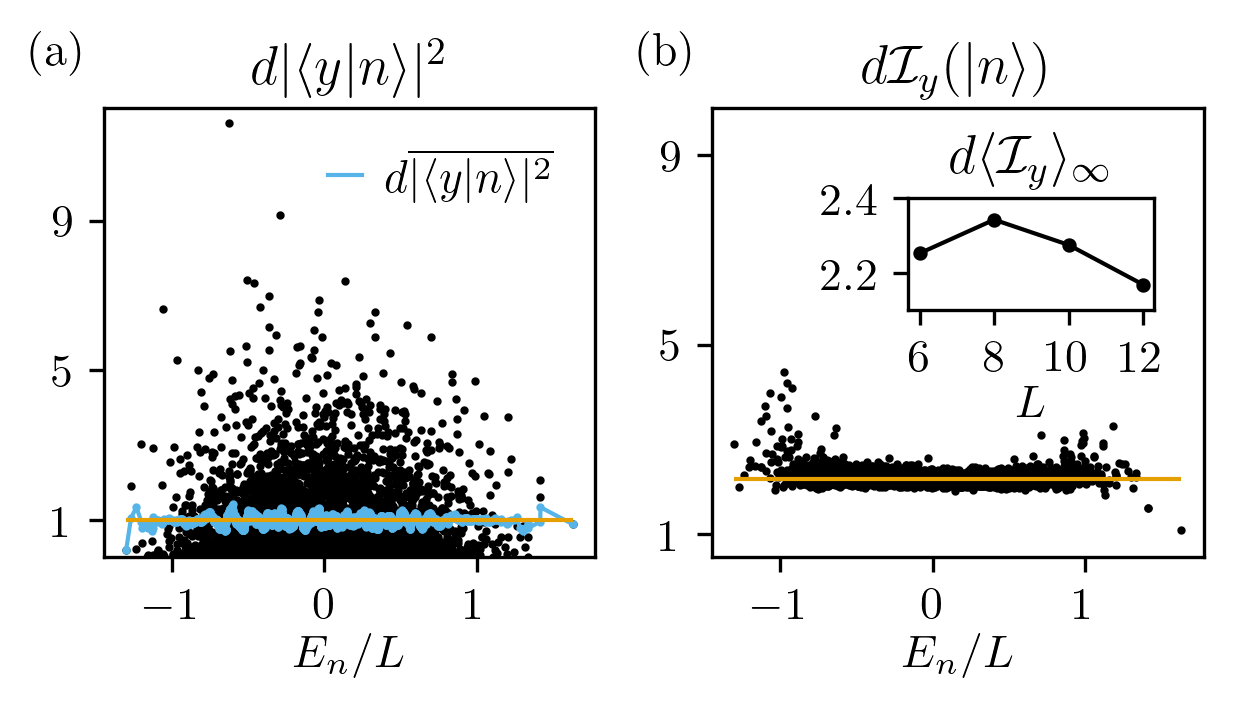}
\centering
\setlength{\abovecaptionskip}{-10pt}
\caption{Relationship between $y$-basis product states and eigenstates of the chaotic MFIM for $L=12$. (a) In black, the overlaps $|\braket{y|n}|^2$ (scaled by $d$) versus energy density for a single random $y$-basis state. In blue, a coarse-grained version of the same. Orange is a horizontal line at $1$. (b) In black, the IPRs $\mathcal{I}_y(\ket{n})$ (scaled by $d$) as a function of energy density. The orange horizontal line represents the spectral averaged IPR $\braket{\mathcal{I}_y}_\infty$. The inset shows that $d \braket{\mathcal{I}_y}_\infty= O(1)$.}
\label{fig:overlaps}
\end{figure}

In Ref.~\cite{hartmann_gaussian_2004} it was shown that the energy distribution of random product states converges in the thermodynamic limit to a normal distribution. In particular, we have for any constants $a<b$ and in the limit of $L\rightarrow\infty$ that
\begin{equation}\label{eq:prod_dist}
    \sum_{a\leq E_n\leq b} |\braket{p|n}|^2 \rightarrow \int_a^b dE \rho_p(E)
\end{equation}
where $\rho_p(E) = (2\pi \sigma^2_p)^{-1/2} e^{-(E-E_p)^2/2\sigma^2_p}$ with $E_p$ and $\sigma^2_p$ the energy and the energy variance, respectively, of the product state $\ket{p}$. In the spirit of Srednicki's formulation of ETH \cite{srednicki_approach_1999} and in a manner consistent with Eq.~\eqref{eq:prod_dist} let us make the phenomenological ansatz
\begin{equation}\label{eq:phenom}
    |\braket{p|n}|^2 = \mathcal{D}^{-1}(E_n)\rho_p(E_n) (1+r_{pn})
\end{equation}
with $-1\leq r_{pn} \leq O(1)$ pseudo-random fluctuations whose particular distribution will not be too important for our purposes. This description is consistent with Eq.~\eqref{eq:prod_dist} since
\begin{multline}
    \sum_{a\leq E_n\leq b} |\braket{p|n}|^2
    =  \sum_{a\leq E_n\leq b} \mathcal{D}^{-1}(E_n)\rho_p(E_n) \\ +
    \sum_{a\leq E_n\leq b} \mathcal{D}^{-1}(E_n) \rho_p(E_n) r_{pn}
\end{multline}
and the first term can be converted to the continuum yielding the limiting form of Eq.~\eqref{eq:prod_dist}. For a chaotic spin chain the density of states is well approximated by a Gaussian,
\begin{equation}
    \mathcal{D}(E) = \sum_n \delta_\epsilon(E-E_n) = d (2\pi \Sigma^2)^{-1/2} e^{-E^2/2\Sigma^2},
\end{equation}
where $\delta_\epsilon$ is a smeared delta function and where we have set the mean of the Gaussian to be $\braket{H}_\infty=0$ and the variance to be
\begin{multline}
    \Sigma^2 = \braket{H^2}_\infty
    =N^2 (L-2)\\ + 2(V^2/4 +h^2_x + h^2_z/4) = O(L) .
\end{multline}
Using this formula and treating $r_{pn}$ as independent and identically distributed random variables with mean zero and higher moments $\mathbb{E}r^k_{pn}=\mu_k$, the second term has zero mean and variance
\begin{multline}
    \sum_{a\leq E_n\leq b} \mu_2 \mathcal{D}^{-2}(E_n) \rho^2_p(E_n) \\
    \leq \frac{\mu_2(2\pi \Sigma^2)^{1/2} \rho_p(E_p)}{d} (b-a)e^{b^2/2\Sigma^2}    
\end{multline}
which is $O(d^{-1})$ assuming $a,b,\mu_2=O(1)$ and $\sigma^2_p = O(L)$, so that the left hand side of Eq.~\eqref{eq:prod_dist} converges to the limiting form with a statistical error of $O(d^{-1/2})$. In case $\ket{p}=\ket{y}$ further simplifications also occur due to the fact that $E_y = \braket{H}_\infty = 0$ and $\sigma^2_y = \braket{H^2}_\infty$. Namely, in this case we have $\mathcal{D}^{-1}(E_n)\rho_y(E_n) = d^{-1}$ for all $n$. This is clearly seen in Fig.~\ref{fig:overlaps}, whereby the coarse grained version of $d|\braket{y|n}|^2$ (obtained by averaging over windows of $64$ nearest eigenenergies), which is an approximation to $d \mathcal{D}^{-1}(E_n)\rho_y(E_n)$, is a very flat function of energy. Still assuming $r_{pn}$ to be mean zero i.i.d. random variables with higher moments $\mathbb{E}r^k_{pn}=\mu_k$, we find
\begin{equation}
    \mathbb{E} \braket{\mathcal{I}_y}_\infty = d^{-1}(1+\mu_2).
\end{equation}

\begin{figure}
\includegraphics[width=\columnwidth]{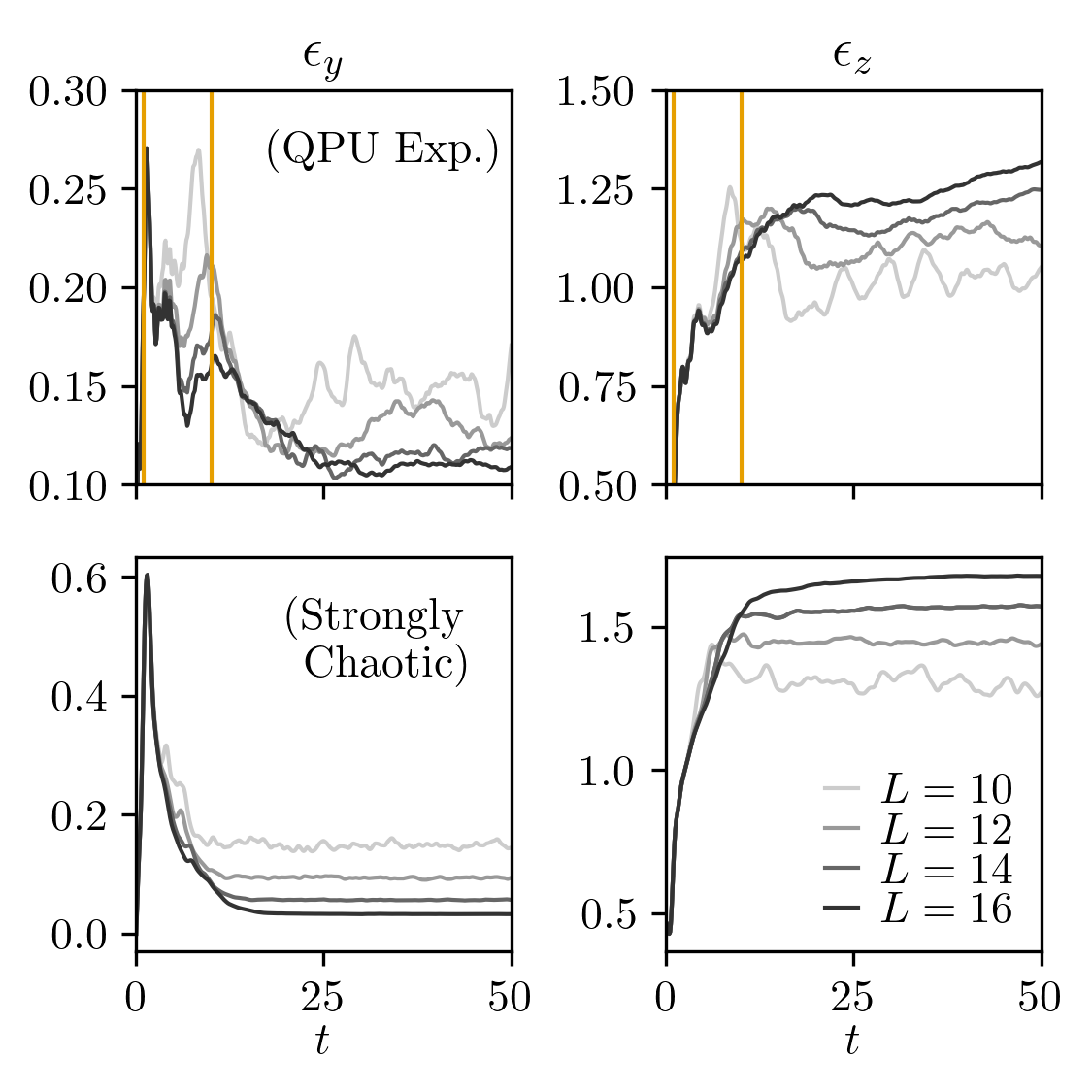}
\centering
\setlength{\abovecaptionskip}{-5pt}
\caption{The dynamical relative error $\epsilon_p$ for product state bases $p=y$ and $p=z$ shown in left and right columns, respectively, for ``QPU experiment" (top) and ``strongly chaotic" (bottom) parameters [see below Eq.~\eqref{eq:h_edge}]. Different curves show data for system sizes $L=10,12,14,16$ with darker gray indicating larger $L$. For the QPU experiment parameters, the region between the orange vertical lines indicate the time range in which the dynamical exponent was extracted in the QPU experiment.}
\label{fig:rel_err}
\end{figure}

We can also calculate how typical this average value is in the joint distribution of $r_{pn}$. We find that the variance is
\begin{multline}
    \mathbb{E} \braket{\mathcal{I}_y}^2_\infty - \big[ \mathbb{E} \braket{\mathcal{I}_y}_\infty\big]^2  \\
    = d^{-4}(4\mu_2 + 4\mu_3 + \mu_4-\mu_2^2).
\end{multline}
Therefore if we assume that $\mu_k$ are at most, say, growing polynomially with $L$ and not exponentially we find for large $L$ that
\begin{equation}
    \braket{\mathcal{I}_y}_\infty = O(d^{-1})
\end{equation}
since statistical fluctuations are exponentially smaller than the average. Interestingly, Fig.~\ref{fig:overlaps} (b) is numerical evidence that at least $\mu_2$ does not grow with $L$ (for the small range of system sizes considered). With a more systematic numerical study of ``how random" the $r_{pn}$ actually are in a nonintegrable system, Eq.~\eqref{eq:phenom} could potentially be seen as part of a more detailed phenomenological understanding of why diagonal ensembles of product states are exponentially small in chaotic systems. For the purposes of this paper, simply assuming that $r_{pn}$ are effectively independent random variables helps justify the central claim of this Appendix; that
\begin{equation}
    \overline{\mathbb{E}_y\big[|\Delta^{ij}_y(t)|^2\big]} \leq O\bigg(\frac{L^{\alpha}}{d}\bigg).
\end{equation}
for some $\alpha>0$ which is consistent with our numerics [recall our earlier discussion of the inset of Fig.~\ref{fig:eth_stuff}(b)].

\subsection{Relative Error}\label{sec:rel_error}

Analytical bounds aside, to further clarify the advantage of sampling $\braket{h_i(t)h_j(0)}_\infty$ in the $y$-basis over some other product state basis, we compare numerically the dynamical relative error
\begin{equation}
    \epsilon_p(t) = \frac{F_0(L,t)}{C_0(t)}
\end{equation}
for the two different product state bases $\ket{p}=\ket{y},\ket{z}$. Fig.~\ref{fig:rel_err} shows a big practical difference between the $y$-basis and $z$-basis. Firstly, the $z$-basis fluctuations in both models are actually larger than the correlator itself at late times and still large at earlier times. On the other hand, in both models the $y$-basis has fluctuations on the order of $20\%$. Consequently, a single randomly drawn $y$-basis state will already approximate the exact correlator with a statistical error of order $20\%$.

\section{Protocol Simplifications made for Hardware Simulations}\label{sec:simplify}

\begin{figure}
\includegraphics[width=\columnwidth]{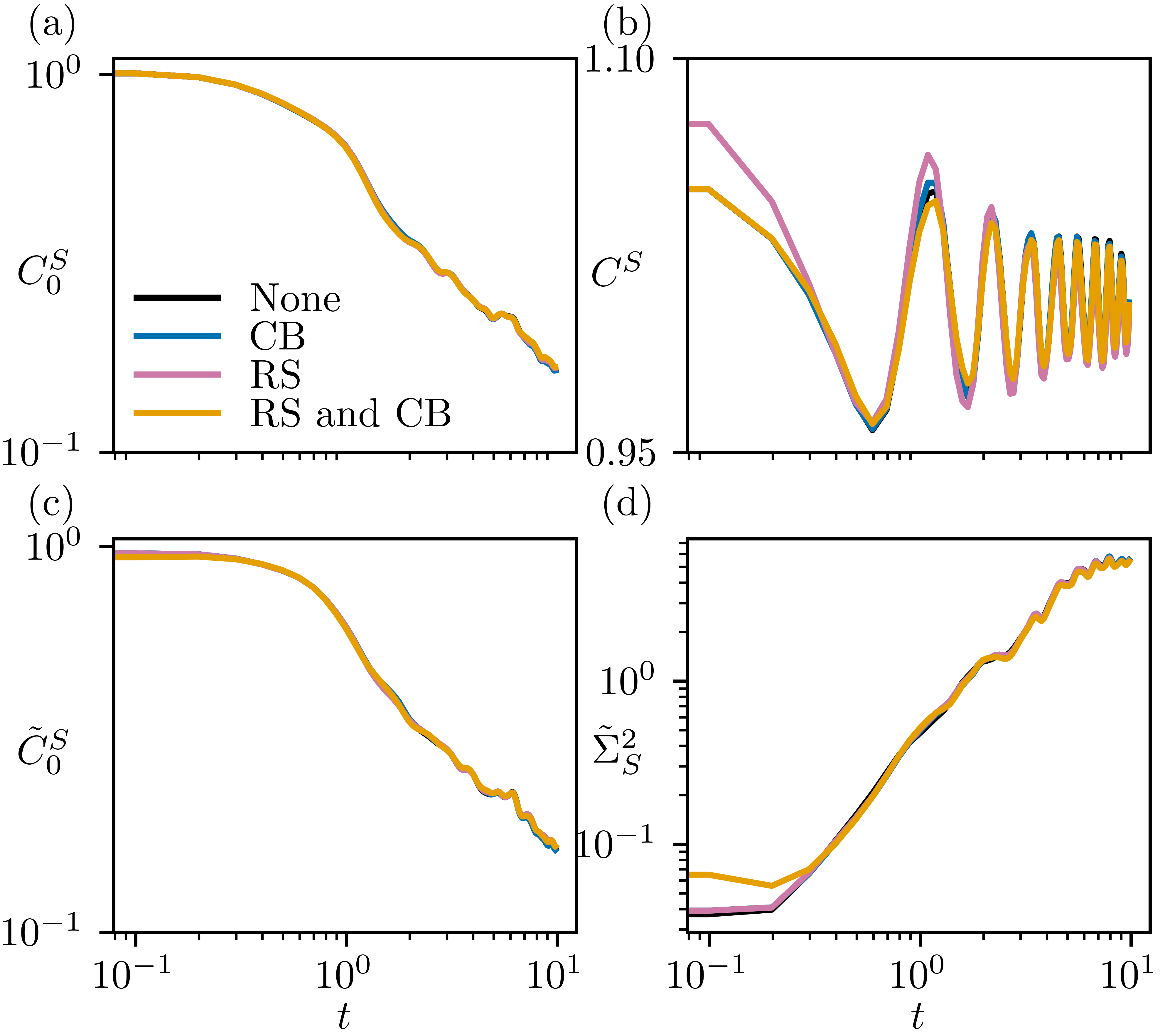}
\centering
\caption{Ideal (no shot noise) Trotter simulation of the effects of making two simplifications to the hardware implementation protocol. (None: without any simplification, CB: replacing Bell pairs with local computational states, RS: assuming reflection symmetry of the ensemble, and CB and RS: combining both simplifications.) All four possible combinations are shown. Model parameters, hyper-parameters, and the set of $y$ basis states [Eq.~\eqref{eq:12_y-sample}] are identical to Fig.~\ref{fig:Om_2} in the main text.}
\label{fig:EX}
\end{figure}

For the hardware simulations, in this appendix, and in the subsequent appendices, we refer to the fixed ensemble of 12 $y$-basis states defined by
\begin{align}
\label{eq:12_y-sample}
\begin{split}
    &100010111110 \quad \quad 010001100101 \\
    &110101111101 \quad \quad 010001111011 \\
    &011101101100 \quad \quad 011101000001 \\
    &100011011010 \quad \quad 111010110010 \\
    &000011010110 \quad \quad 111110001111 \\
    &001011001110 \quad \quad 011011101000
\end{split}
\end{align}
where $1$ denotes a $+1$ eigenstate of $Y$ on the corresponding site and $0$ denotes a $-1$ eigenstate. 

\begin{figure*}[t]
    \centering
    \includegraphics[width=1.8\columnwidth]{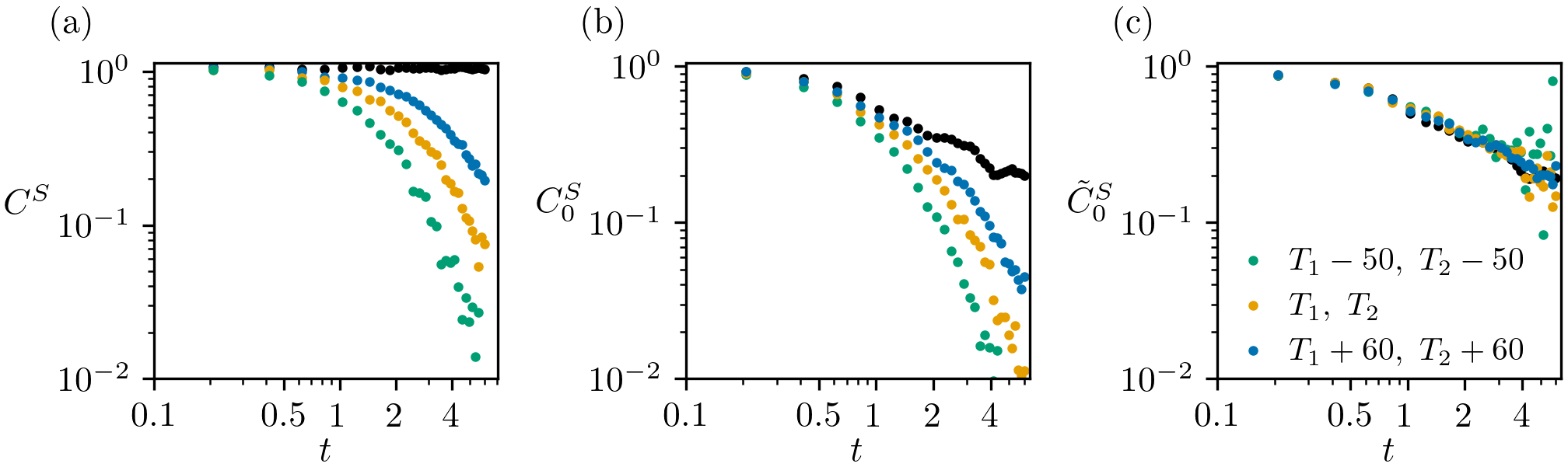}
    \caption{Sum rule and the energy autocorrelators from the noisy simulator with $8192$ shots. All parameters and hyper-parameters are the same as in Fig.~\ref{fig:Om_2}, including the choice of $y$-basis sample [Eq.~\eqref{eq:12_y-sample}]. (a) The sum rule results with different thermal relaxation time set $(T_{1}=120.73\mu \text{s},T_{2}=107.29\mu \text{s})$, $\left(T_{1}-50\mu \text{s},T_{2}-50\mu \text{s}\right)$, and $\left(T_{1}+60\mu \text{s},T_{2}+60\mu \text{s}\right)$. (b) The bare sample-averaged energy autocorrelator. (c) The renormalized sample-averaged energy autocorrelator. The black points represent the results of a noiseless circuit simulation of the same Trotter dynamics.} 
    \label{fig:noise}
\end{figure*}

In Sec.~\ref{Sec:QC} of the main text, we discussed two principle simplifying assumptions we make in implementing the transport experiment on hardware. We assume reflection symmetry in Eq.~\ref{eq:ensemble_symmetry} in order to reduce the number of circuits. Besides that, we also evolve local computational basis states $\{\ket{00},\ket{01},\ket{10},\ket{11}\}$ instead of the local Bell states $\ket{\pm,\nu,y}$ (when $\nu=3,4$) in order to reduce circuit depth. The error made in replacing the Bell states present in, for example, the term
\begin{equation}
    \braket{+,3,y|P^\mu_{L/2+r}(t)|+,3,y}
\end{equation}
with product states comes from what we call the ``off diagonal" contribution
\begin{align}
 (-1)^{y_{\frac{L}{2}}+y_{\frac{L}{2}+1}}\braket{P^\mu_{L/2+r}(t)}^y_{\text{off}}
 \label{eq:diff}
\end{align}
where $y_j$ denotes $j$th entry of a bitstring in the list \eqref{eq:12_y-sample} and
\begin{multline}
     4 \braket{P^\mu_{L/2+r}(t)}^y_{\text{off}}= \\
     - \braket{00,y|P^\mu_{L/2+r}(t)|11,y} +\braket{11,y|P^\mu_{L/2+r}(t)|00,y} \\
     + \braket{01,y|P^\mu_{L/2+r}(t)|10,y} - \braket{10,y|P^\mu_{L/2+r}(t)|01,y}.
\end{multline}
with $\ket{00,y}, \ket{01,y}, \ket{10,y}, \ket{11,y}$ denoting product states obtained by replacing the $Y$ eigenstates on sites $L/2-1$ and $L/2$ with the corresponding $z$-basis states indicated by the first argument. Eq.~\ref{eq:diff} depends explicitly on the $y$-basis sample being averaged over. Under an average over $S$ samples, Eq.~\eqref{eq:diff} will be suppressed by a prefactor $S^{-1/2}$ due to the random sign. However, the smallness of Eq.~\ref{eq:diff} for relatively small $S$ depends on the size of $\braket{P^\mu_{L/2+r}(t)}^y_{\text{off}}$.

Fig.~\ref{fig:EX} shows statevector simulator results (i.e., assuming an infinite number of measurement shots) for Trotter simulations of the sample-averaged energy correlation function and SV using just Eq.~\ref{eq:mitarai_fuji} (None), using computational basis states instead of Bell states so as to ignore Eq.~\eqref{eq:diff} (CB), assuming reflection symmetry (RS), and combining both simplifications (CB and RS). 
For all sample-averaged quantities considered, namely the bare correlator $C^S_0$ (a), the sum rule value $C^S$ (b), the renormalized correlator $\tilde C^S_0$ (c), and the corresponding SV (d), all four methods yield nearly indistinguishable results. This indicates that our simplifying assumptions should not have a strong impact on the QPU results.

\section{Noise Analysis}
\label{sec:noise}

In this appendix we simulate the impact of noise on the sum rule value $C(t)=\sum_rC_r(t)$ and the renormalization strategy discussed in the main text that corrects for the noise-induced decay of this quantity with time.
Motivated by the fact that two-qubit gates take significantly longer to perform on hardware than one-qubit gates, our noisy circuit simulation in Qiskit adds noise only to the two-qubit gates in the Trotter circuit.
We model thermal relaxation noise during the application of a two-qubit gate by subsequently acting with a phenomenological noise channel including both dephasing and amplitude damping components~\cite{Joydip12}:
\begin{align}
    \mathcal{N}_{th}(\rho)=\sum_{i}V_{i}\rho V_{i}^{\dagger}, 
\end{align}
with Kraus operators
 \begin{align}
    V_{0}=\left(\begin{array}{cc}
1 & 0\\
0 & e^{-i\delta\omega t}e^{-t/T_{2}}
\end{array}\right)
\end{align}
 \begin{align}
V_{1}=\left(\begin{array}{cc}
0 & 0\\
0 & \sqrt{e^{-t/T_{1}}-e^{-t/T_{2}}}
\end{array}\right)
\end{align}
 \begin{align}
V_{2}=\left(\begin{array}{cc}
0 & \sqrt{1-e^{-t/T_{1}}}\\
0 & 0
\end{array}\right).
\end{align}
Here, $T_1$ and $T_2$ are the amplitude damping and dephasing relaxation times, respectively, $t$ is the duration of the two-qubit gate, and $\delta\omega$ is the energy difference between $\ket{0}$ and $\ket{1}$. We fix $t=0.6\ \mu s$ and test three different parameter sets $(T_{1}=120.7\ \mu s,T_{2}=107.3\ \mu s)$, $\left(T_{1}-50,T_{2}-50\right)$, and $\left(T_{1}+60,T_{2}+60\right)$.
The parameters $t$, $T_1$, and $T_2$ are chosen to be comparable to those of real IBM QPUs, which on average have roughly $T_1 \approx 100 \mu s$ and $T_2 \approx 100 \mu s$.

We perform noisy circuit simulations in Qiskit by applying the above noise channel after each two-qubit gate to the qubits on which the gate acted. 
Results for the sum rule and energy density autocorrelators are shown in Fig.\ref{fig:noise} [with results averaged over the same $y$-basis sample as in the QPU experiment, i.e., Eq.~\eqref{eq:12_y-sample}].
Fig.\ref{fig:noise} (a) shows that the sum rule decays faster as $T_1$ and $T_2$ become shorter.
This decay impacts the energy density autocorrelator results shown in panel (b) in a predictable manner---the global nonconservation of energy due to noise leads to faster decay of the autocorrelator, with the decay rate of the sum rule correlating directly with the deviation from the results of a noiseless circuit simulation (black points).
In panel (c) we see that the renormalized correlators in the presence of noise match the exact version much better. The timescale on which visible deviations develop can be directly inferred from panel (a): once the sum rule decays to a value $\sim 0.2$ the renormalized correlator becomes noisy due to the smallness of both the numerator and denominator of Eq.~\eqref{eq:Ctilde_def}. This demonstrates how the renormalization strategy adopted in the main text can be used to systematically reduce the impact of noise.

\section{Additional QPU Data}
\label{sec:OtherParams}

\begin{figure}
    \centering
    \includegraphics[width=1.0\columnwidth]{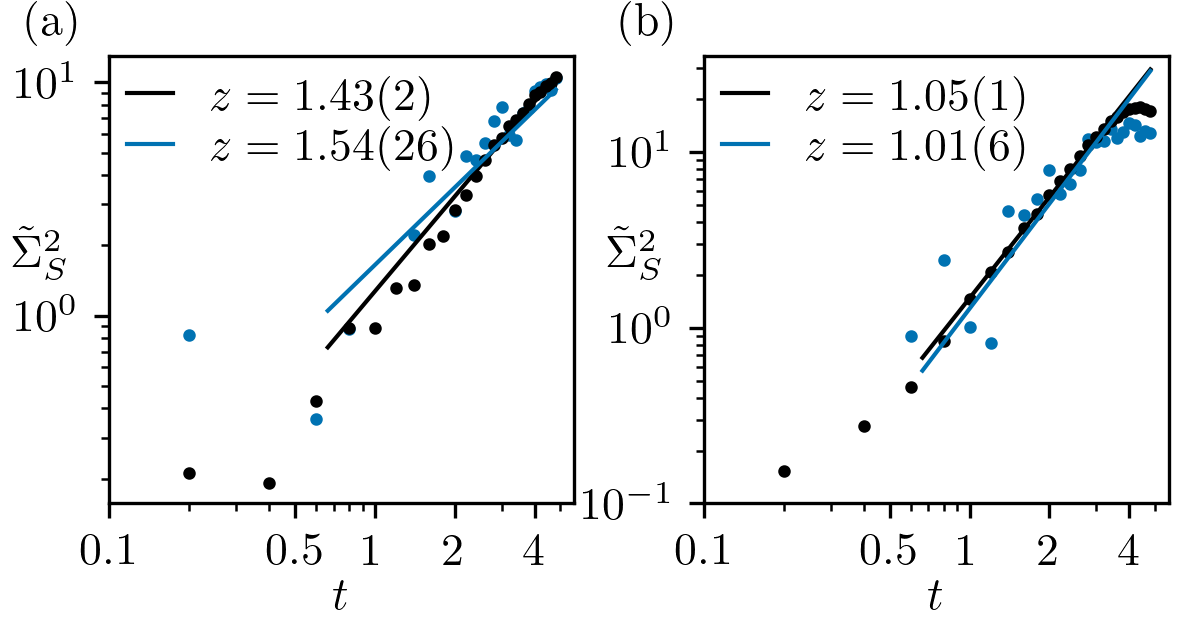}
    \caption{Renormalized spatial variance with different parameters for (a) $\Omega=3$ and (b) $\Omega=6$ from \texttt{ibmq\_montreal}. Remaining parameters and hyper-parameters are as in Fig.~\ref{fig:Om_2}. Blue and black points are data from the QPU and ideal simulator, respectively. Blue and black lines depict power-law fits to $bt^{2/z}$ for the QPU and simulator data, respectively.}
    \label{fig:s_other}
\end{figure}

\begin{figure}
\includegraphics[width=1.0\columnwidth]{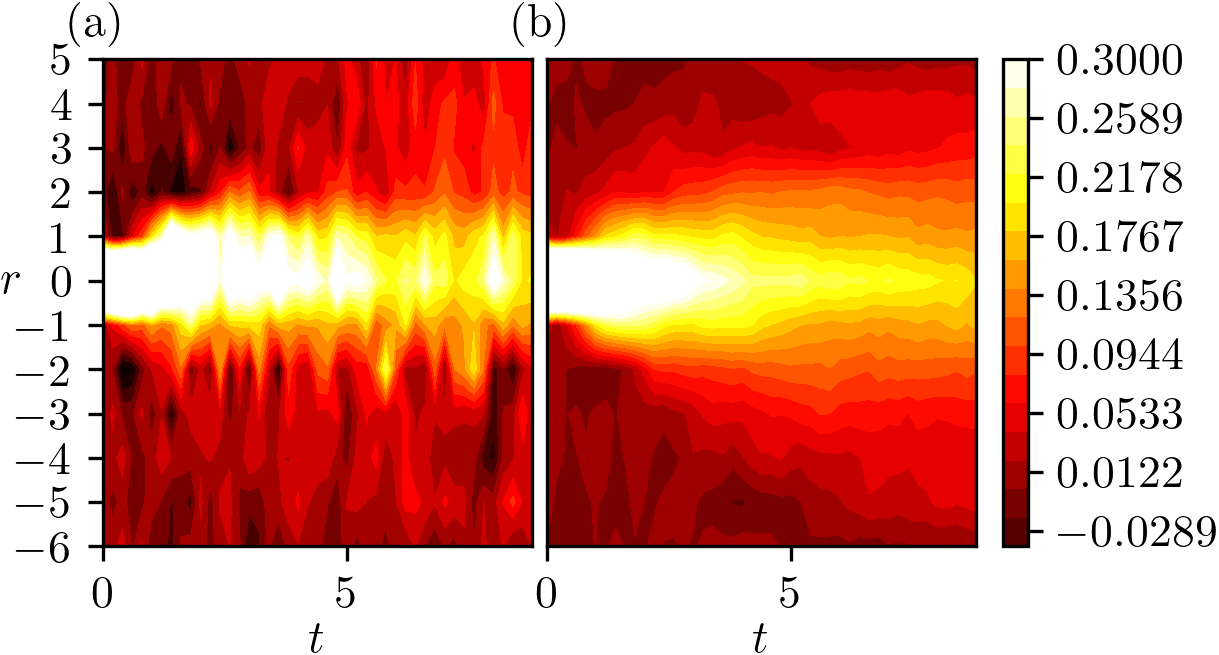}
\centering
\caption{Spatio-temporal visualization of the renormalized energy correlation functions $\tilde{C}^S_r$ obtained via (a) hardware experiment on \texttt{ibmq\_montreal} and (b) ideal simulator. Model parameters and hyper-parameters are identical to those in Fig.~\ref{fig:Om_2}.}
\label{fig:Om_2space}
\end{figure}

\begin{figure}
    \centering
    \includegraphics[width=1.0\columnwidth]{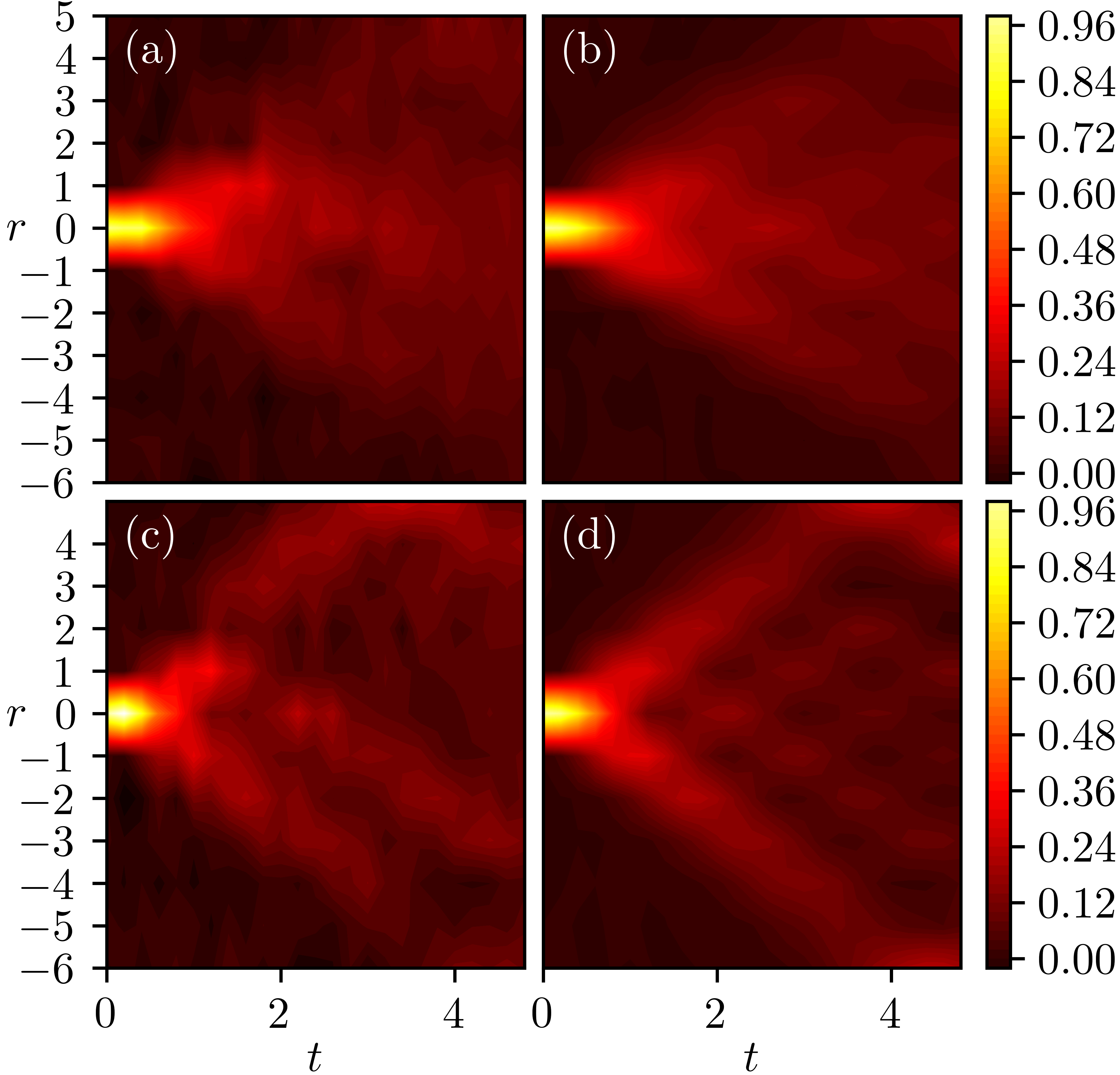}
    \caption{Renormalized energy correlators $\tilde C^S_r$ with $\Omega=3$ from (a) QPU and (b) ideal simulator, and with $\Omega=6$ from (c) QPU and (d) ideal simulator. Remaining model parameters and all hyper-parameters are identical to those in Fig.~\ref{fig:Om_2}. }
    \label{fig:C_other}
\end{figure}

In the main text we showed in Fig.~\ref{fig:others} that the energy autocorrelation functions with $\Omega=3$ and $\Omega=6$ exhibit superdiffusive ($\tilde{C}_{0}^{S}(t) \propto t^{-0.645}$) and ballistic ($\tilde{C}_{0}^{S}(t) \propto t^{-1.036}$) scaling at early times, respectively. In Fig.~\ref{fig:s_other} we cross-reference QPU and simulator results for the renormalized SV. 
For both choices of $\Omega$, the power-law growth exponents match well between the QPU and simulator results (blue and black points, respectively). 
Furthermore, hydrodynamics predict that $\Sigma^2\sim t^{2/z}$ when $C_0\sim t^{-1/z}$. 
The exponents extracted from power-law fits to the QPU results (see plot legends) are consistent with this prediction, being very close to twice the decay exponents extracted from the results in Fig.~\ref{fig:others}.

Consistent with the energy density autocorrelator results in Fig.~\ref{fig:others}, the SV results in Fig.~\ref{fig:s_other} display good agreement between the QPU and the simulator. 
This reinforces the point made in the main text that the faster transport timescale for larger $\Omega$ leaves less time for decoherence to affect the QPU results.

Figs.~\ref{fig:Om_2space} and \ref{fig:C_other} show the spatial profile of the renormalized energy density correlators $\tilde C^S_r$ for $\Omega=2$ and $\Omega=3,6$, respectively. The results shown in Fig.~\ref{fig:Om_2space} are much noisier than those shown in Fig.~\ref{fig:C_other}, consistent with the fact that the $\Omega=2$ QPU simulation is more challenging due to the longer evolution time necessary to capture the decay of the autocorrelator. In particular, while an approximately diffusive $\sim\sqrt{t}$ light cone is visible in the circuit simulation in Fig.~\ref{fig:Om_2space}(b), no such feature is visible for the QPU results in Fig.~\ref{fig:Om_2space}(a) due to noise. However, for larger $\Omega$, where the noise has less time to impact the results, the QPU and simulator results are qualitatively very similar, see Fig.~\ref{fig:C_other}. In particular, the energy wavefront or light cone structure is clearly visible in both the QPU and simulator plots. Interestingly, the results in Fig.~\ref{fig:C_other} also reveal spatial variations in the quality of individual qubits. For example, in Fig.~\ref{fig:C_other}(c) there is a dark spot appearing at site $r=2$ near $t=1.6$ that does not appear in the simulator results from Fig.~\ref{fig:C_other}(d). Indeed, the corresponding qubit has lower $T_1$ and $T_2$. 

\section{Haar Random Product States}\label{sec:haar_product}

In this appendix we study the problem of using products of Haar random single-qudit states for sampling the trace of an operator $Q$. We (1) derive an exact expression for the sample-to-sample fluctuations in estimating $\braket{Q}_\infty$ in this ensemble and (2) show that this quantity is bounded by $\braket{Q^\dagger Q}_\infty$. 

Take $Q$ to act non-trivially on $N$ sites of a system with local Hilbert space dimension $q$. Consider the random ensemble of states  $\ket{\psi}=\ket{\phi_1}\cdots \ket{\phi_N}$ where $\ket{\phi_k}$ are i.i.d. Haar random states acting on one site, i.e. each realized by $U \ket{0}$ for $U$ a $q\times q$ Haar random unitary and $\ket{0}$ an arbitrary fixed state. First, it is clear that these states are unbiased estimators of the trace, i.e.
\begin{equation}
    \mathbb{E} \ \braket{\psi|Q|\psi} = \braket{Q}_\infty.
\end{equation}
We claim that the second moment $\mathbb{E}|\braket{\psi|Q|\psi}|^2 $ is given exactly by
\begin{equation}\label{eq:haar_prod_fluct}
    \frac{1}{(q(q+1))^{N}}  \sum_{x\in\{0,1\}^N}  \text{tr}_{\bar{x}} (\text{tr}_x(Q^\dagger)\text{tr}_x(Q)).
\end{equation}
\normalsize
Here $\text{tr}_x$ means to trace all sites $j$ out of the total $N$ for which $x_j=1$ and $\bar{x}$ is the logical NOT of $x$. We sketch the proof of this as follows. The central formula is the correlator
\begin{multline}
    \mathbb{E}_{U}[\braket{0|U^\dagger|i} \braket{j|U|0} \braket{0|U^\dagger|k} \braket{l|U|0}]  \\ = \frac{1}{q(q+1)}(\delta_{ij}\delta_{kl} + \delta_{il}\delta_{kj})
\end{multline}
which involves only the first column of a Haar random $U$ and follows from the expression for the general four point correlator of a Haar matrix \cite{collins_integration_2006}. In a product basis of the $N$ sites, we find
\begin{multline}
    (q(q+1))^{-N} \sum_{IJKL} \braket{I|Q^\dagger|J}\braket{K|Q|L} \\
    \times \prod_{s=1}^N (\delta_{i_sj_s}\delta_{k_sl_s} + \delta_{i_sl_s}\delta_{k_sj_s})
\end{multline}
where $I = i_1 \cdots i_s\cdots i_N$ etc. are multi-indices. Expanding out the product of delta functions leads to $2^N$ possibilities, where in each case, some of the sub-indices in $I$ are connected to those in $J$ and some are connected to those in $L$. Each time a sub-index in $I$ is connected to one in $J$, a partial trace is induced. The situation can be summarized as Eq.~\eqref{eq:haar_prod_fluct}.

One can now obtain from this formula an upper bound. First observe that each term inside the sum is a square Frobenius norm of the matrix $\text{tr}_x(Q)$. Due to Rastegin \cite{rastegin_relations_2012}, the norm of this partial trace can be related to the Frobenius norm of the full matrix $Q$ via $||\text{tr}_x(Q)||_F \leq \sqrt{D} ||Q||_F$ where $D$ is the dimension of the space being traced out. In this case $D = q^{w(x)}$ where $w(x)$ is the Hamming weight (number of ones present) in string $x$. Putting this in and using the binomial theorem we obtain the bound
\begin{multline}
    \mathbb{E} \ |\braket{\psi|Q|\psi}|^2 \\
    \leq ||Q||^2_F\ (q(q+1))^{-N} \sum_{x\in\{0,1\}^N} q^{w(x)} = \braket{Q^\dagger Q}_\infty.
\end{multline}
The actual sample-to-sample fluctuations are controlled by the variance $|\braket{\psi|Q|\psi}|^2 - |\braket{Q}|^2_\infty$, but in principle $\braket{Q^\dagger Q}_\infty$ still provides a rigorous upper bound.
\end{appendix}

\bibliographystyle{quantum}
\bibliography{main}

\begin{thebibliography}{10}

\bibitem{Doyon20}
Benjamin Doyon.
\newblock ``{Lecture notes on Generalised Hydrodynamics}''.
\newblock \href{https://dx.doi.org/10.21468/SciPostPhysLectNotes.18}{SciPost
  Phys. Lect. NotesPage~18}~(2020).

\bibitem{Schollwock11}
Ulrich Schollw{\"o}ck.
\newblock ``The density-matrix renormalization group in the age of matrix
  product states''.
\newblock
  \href{https://dx.doi.org/10.1016/j.aop.2010.09.012}{Ann.~Phys.~(N.~Y.) {\bf
  326}, 96--192}~(2011).

\bibitem{Orus2014}
Rom{\'{a}}n Or{\'{u}}s.
\newblock ``A practical introduction to tensor networks: Matrix product states
  and projected entangled pair states''.
\newblock \href{https://dx.doi.org/10.1016/j.aop.2014.06.013}{Annals of Physics
  {\bf 349}, 117--158}~(2014).

\bibitem{Bertini2021}
B.~Bertini, F.~Heidrich-Meisner, C.~Karrasch, T.~Prosen, R.~Steinigeweg, and
  M.~\ifmmode \check{Z}\else \v{Z}\fi{}nidari\ifmmode~\check{c}\else
  \v{c}\fi{}.
\newblock ``Finite-temperature transport in one-dimensional quantum lattice
  models''.
\newblock \href{https://dx.doi.org/10.1103/RevModPhys.93.025003}{Rev. Mod.
  Phys. {\bf 93}, 025003}~(2021).

\bibitem{Znidaric16}
Marko \ifmmode \check{Z}\else \v{Z}\fi{}nidari\ifmmode~\check{c}\else
  \v{c}\fi{}, Antonello Scardicchio, and Vipin~Kerala Varma.
\newblock ``Diffusive and subdiffusive spin transport in the ergodic phase of a
  many-body localizable system''.
\newblock \href{https://dx.doi.org/10.1103/PhysRevLett.117.040601}{Phys. Rev.
  Lett. {\bf 117}, 040601}~(2016).

\bibitem{Basko06}
D.~M. Basko, I.~L. Aleiner, and B.~L. Altshuler.
\newblock ``Metal{\textendash}insulator transition in a weakly interacting
  many-electron system with localized single-particle states''.
\newblock
  \href{https://dx.doi.org/10.1016/j.aop.2005.11.014}{Ann.~Phys.~(N.~Y.) {\bf
  321}, 1126--1205}~(2006).

\bibitem{Gornyi-PRL-2005}
I.~V. Gornyi, A.~D. Mirlin, and D.~G. Polyakov.
\newblock ``Interacting electrons in disordered wires: Anderson localization
  and low-$t$ transport''.
\newblock \href{https://dx.doi.org/10.1103/PhysRevLett.95.206603}{Phys. Rev.
  Lett. {\bf 95}, 206603}~(2005).

\bibitem{Yu16}
Xiongjie Yu, David~J. Luitz, and Bryan~K. Clark.
\newblock ``Bimodal entanglement entropy distribution in the many-body
  localization transition''.
\newblock \href{https://dx.doi.org/10.1103/PhysRevB.94.184202}{Phys. Rev. B
  {\bf 94}, 184202}~(2016).

\bibitem{schulzEnergyTransportDisordered2018}
M.~Schulz, S.~R. Taylor, C.~A. Hooley, and A.~Scardicchio.
\newblock ``Energy transport in a disordered spin chain with broken {{U}}(1)
  symmetry: {{Diffusion}}, subdiffusion, and many-body localization''.
\newblock \href{https://dx.doi.org/10.1103/PhysRevB.98.180201}{Phys. Rev. B
  {\bf 98}, 180201}~(2018).

\bibitem{doggenManybodyLocalizationLarge2021}
Elmer V.~H. Doggen, Igor~V. Gornyi, Alexander~D. Mirlin, and Dmitry~G.
  Polyakov.
\newblock ``Many-body localization in large systems: {{Matrix-product-state}}
  approach''.
\newblock \href{https://dx.doi.org/10.1016/j.aop.2021.168437}{Annals of Physics
  {\bf 435}, 168437}~(2021).

\bibitem{feldmeier_anomalous_2020}
Johannes Feldmeier, Pablo Sala, Giuseppe De~Tomasi, Frank Pollmann, and Michael
  Knap.
\newblock ``Anomalous {Diffusion} in {Dipole}- and
  {Higher}-{Moment}-{Conserving} {Systems}''.
\newblock \href{https://dx.doi.org/10.1103/PhysRevLett.125.245303}{Physical
  Review Letters {\bf 125}, 245303}~(2020).

\bibitem{Wei22}
David Wei, Antonio Rubio-Abadal, Bingtian Ye, Francisco Machado, Jack Kemp,
  Kritsana Srakaew, Simon Hollerith, Jun Rui, Sarang Gopalakrishnan, Norman~Y.
  Yao, Immanuel Bloch, and Johannes Zeiher.
\newblock ``Quantum gas microscopy of kardar-parisi-zhang superdiffusion''.
\newblock \href{https://dx.doi.org/10.1126/science.abk2397}{Science {\bf 376},
  716--720}~(2022).

\bibitem{Keenan22}
Nathan Keenan, Niall Robertson, Tara Murphy, Sergiy Zhuk, and John Goold.
\newblock ``Evidence of kardar-parisi-zhang scaling on a digital quantum
  simulator''.
\newblock \href{https://dx.doi.org/10.1038/s41534-023-00742-4}{npj Quantum
  Information{\bf 9}}~(2023).

\bibitem{Gopalakrishnan23}
Sarang Gopalakrishnan and Romain Vasseur.
\newblock ``Anomalous transport from hot quasiparticles in interacting spin
  chains''.
\newblock \href{https://dx.doi.org/10.1088/1361-6633/acb36e}{Reports on
  Progress in Physics {\bf 86}, 036502}~(2023).

\bibitem{Ljubotina23}
Marko Ljubotina, Jean-Yves Desaules, Maksym Serbyn, and Zlatko
  Papi\ifmmode~\acute{c}\else \'{c}\fi{}.
\newblock ``Superdiffusive energy transport in kinetically constrained
  models''.
\newblock \href{https://dx.doi.org/10.1103/PhysRevX.13.011033}{Phys. Rev. X
  {\bf 13}, 011033}~(2023).

\bibitem{Haegeman11}
Jutho Haegeman, J.~Ignacio Cirac, Tobias~J. Osborne, Iztok
  Pi\ifmmode~\check{z}\else \v{z}\fi{}orn, Henri Verschelde, and Frank
  Verstraete.
\newblock ``Time-dependent variational principle for quantum lattices''.
\newblock \href{https://dx.doi.org/10.1103/PhysRevLett.107.070601}{Phys. Rev.
  Lett. {\bf 107}, 070601}~(2011).

\bibitem{Haegeman16}
Jutho Haegeman, Christian Lubich, Ivan Oseledets, Bart Vandereycken, and Frank
  Verstraete.
\newblock ``Unifying time evolution and optimization with matrix product
  states''.
\newblock \href{https://dx.doi.org/10.1103/PhysRevB.94.165116}{Phys. Rev. B
  {\bf 94}, 165116}~(2016).

\bibitem{Leviatan17}
Eyal Leviatan, Frank Pollmann, Jens~H. Bardarson, David~A. Huse, and Ehud
  Altman.
\newblock ``Quantum thermalization dynamics with matrix-product states''~(2017)
  \href{http://arxiv.org/abs/1702.08894}{arXiv:1702.08894}.

\bibitem{White18}
Christopher~David White, Michael Zaletel, Roger S.~K. Mong, and Gil Refael.
\newblock ``Quantum dynamics of thermalizing systems''.
\newblock \href{https://dx.doi.org/10.1103/PhysRevB.97.035127}{Phys. Rev. B
  {\bf 97}, 035127}~(2018).

\bibitem{Kloss18}
Benedikt Kloss, Yevgeny~Bar Lev, and David Reichman.
\newblock ``Time-dependent variational principle in matrix-product state
  manifolds: Pitfalls and potential''.
\newblock \href{https://dx.doi.org/10.1103/PhysRevB.97.024307}{Phys. Rev. B
  {\bf 97}, 024307}~(2018).

\bibitem{Rakovszky22}
Tibor Rakovszky, C.~W. von Keyserlingk, and Frank Pollmann.
\newblock ``Dissipation-assisted operator evolution method for capturing
  hydrodynamic transport''.
\newblock \href{https://dx.doi.org/10.1103/PhysRevB.105.075131}{Phys. Rev. B
  {\bf 105}, 075131}~(2022).

\bibitem{Yoo23}
Yongchan Yoo, Christopher~David White, and Brian Swingle.
\newblock ``Open-system spin transport and operator weight dissipation in spin
  chains''.
\newblock \href{https://dx.doi.org/10.1103/PhysRevB.107.115118}{Phys. Rev. B
  {\bf 107}, 115118}~(2023).

\bibitem{Richter21}
Jonas Richter and Arijeet Pal.
\newblock ``Simulating hydrodynamics on noisy intermediate-scale quantum
  devices with random circuits''.
\newblock \href{https://dx.doi.org/10.1103/PhysRevLett.126.230501}{Phys. Rev.
  Lett. {\bf 126}, 230501}~(2021).

\bibitem{Preskill18}
John Preskill.
\newblock ``Quantum computing in the nisq era and beyond''.
\newblock \href{https://dx.doi.org/10.22331/q-2018-08-06-79}{Quantum {\bf 2},
  79}~(2018).

\bibitem{Lorenza98}
Lorenza Viola and Seth Lloyd.
\newblock ``Dynamical suppression of decoherence in two-state quantum
  systems''.
\newblock \href{https://dx.doi.org/10.1103/PhysRevA.58.2733}{Phys. Rev. A {\bf
  58}, 2733--2744}~(1998).

\bibitem{Pokharel18}
Bibek Pokharel, Namit Anand, Benjamin Fortman, and Daniel~A. Lidar.
\newblock ``Demonstration of fidelity improvement using dynamical decoupling
  with superconducting qubits''.
\newblock \href{https://dx.doi.org/10.1103/PhysRevLett.121.220502}{Phys. Rev.
  Lett. {\bf 121}, 220502}~(2018).

\bibitem{Jurcevic2021}
Petar Jurcevic, Ali Javadi-Abhari, Lev~S Bishop, Isaac Lauer, Daniela~F
  Bogorin, Markus Brink, Lauren Capelluto, Oktay Günlük, Toshinari Itoko,
  Naoki Kanazawa, Abhinav Kandala, George~A Keefe, Kevin Krsulich, William
  Landers, Eric~P Lewandowski, Douglas~T McClure, Giacomo Nannicini, Adinath
  Narasgond, Hasan~M Nayfeh, Emily Pritchett, Mary~Beth Rothwell, Srikanth
  Srinivasan, Neereja Sundaresan, Cindy Wang, Ken~X Wei, Christopher~J Wood,
  Jeng-Bang Yau, Eric~J Zhang, Oliver~E Dial, Jerry~M Chow, and Jay~M Gambetta.
\newblock ``Demonstration of quantum volume 64 on a superconducting quantum
  computing system''.
\newblock \href{https://dx.doi.org/10.1088/2058-9565/abe519}{Quantum Science
  and Technology {\bf 6}, 025020}~(2021).

\bibitem{Stenger2021}
John P.~T. Stenger, Nicholas~T. Bronn, Daniel~J. Egger, and David Pekker.
\newblock ``Simulating the dynamics of braiding of majorana zero modes using an
  ibm quantum computer''.
\newblock \href{https://dx.doi.org/10.1103/PhysRevResearch.3.033171}{Phys. Rev.
  Research {\bf 3}, 033171}~(2021).

\bibitem{Kim21}
Youngseok Kim, Christopher~J. Wood, Theodore~J. Yoder, Seth~T. Merkel, Jay~M.
  Gambetta, Kristan Temme, and Abhinav Kandala.
\newblock ``Scalable error mitigation for noisy quantum circuits produces
  competitive expectation values''.
\newblock \href{https://dx.doi.org/10.1038/s41567-022-01914-3}{Nature Physics
  {\bf 19}, 752–759}~(2023).

\bibitem{caiQuantumErrorMitigation2022b}
Zhenyu Cai, Ryan Babbush, Simon~C. Benjamin, Suguru Endo, William~J. Huggins,
  Ying Li, Jarrod~R. McClean, and Thomas~E. O'Brien.
\newblock ``Quantum error mitigation''.
\newblock \href{https://dx.doi.org/10.1103/RevModPhys.95.045005}{Rev. Mod.
  Phys. {\bf 95}, 045005}~(2023).

\bibitem{Li17}
Ying Li and Simon~C. Benjamin.
\newblock ``Efficient variational quantum simulator incorporating active error
  minimization''.
\newblock \href{https://dx.doi.org/10.1103/PhysRevX.7.021050}{Phys. Rev. X {\bf
  7}, 021050}~(2017).

\bibitem{Temme17}
Kristan Temme, Sergey Bravyi, and Jay~M. Gambetta.
\newblock ``Error mitigation for short-depth quantum circuits''.
\newblock \href{https://dx.doi.org/10.1103/PhysRevLett.119.180509}{Phys. Rev.
  Lett. {\bf 119}, 180509}~(2017).

\bibitem{Kandala19}
Abhinav Kandala, Kristan Temme, Antonio~D. Córcoles, Antonio Mezzacapo,
  Jerry~M. Chow, and Jay~M. Gambetta.
\newblock ``Error mitigation extends the computational reach of a noisy quantum
  processor''.
\newblock \href{https://dx.doi.org/10.1038/s41586-019-1040-7}{Nature {\bf 567},
  491–495}~(2019).

\bibitem{Giurgica-Tiron-ZNE-2020}
Tudor Giurgica-Tiron, Yousef Hindy, Ryan LaRose, Andrea Mari, and William~J.
  Zeng.
\newblock ``Digital zero noise extrapolation for quantum error mitigation''.
\newblock In 2020 IEEE International Conference on Quantum Computing and
  Engineering (QCE).
\newblock \href{https://dx.doi.org/10.1109/QCE49297.2020.00045}{Pages
  306--316}.
\newblock ~(2020).

\bibitem{berthusenQuantumDynamicsSimulations2022a}
Noah~F. Berthusen, Tha{\'i}s~V. Trevisan, Thomas Iadecola, and Peter~P. Orth.
\newblock ``Quantum dynamics simulations beyond the coherence time on noisy
  intermediate-scale quantum hardware by variational {{Trotter}} compression''.
\newblock \href{https://dx.doi.org/10.1103/PhysRevResearch.4.023097}{Phys. Rev.
  Res. {\bf 4}, 023097}~(2022).

\bibitem{kimEvidenceUtilityQuantum2023}
Youngseok Kim, Andrew Eddins, Sajant Anand, Ken~Xuan Wei, Ewout {van den Berg},
  Sami Rosenblatt, Hasan Nayfeh, Yantao Wu, Michael Zaletel, Kristan Temme, and
  Abhinav Kandala.
\newblock ``Evidence for the utility of quantum computing before fault
  tolerance''.
\newblock \href{https://dx.doi.org/10.1038/s41586-023-06096-3}{Nature {\bf
  618}, 500--505}~(2023).

\bibitem{Suguru18}
Suguru Endo, Simon~C. Benjamin, and Ying Li.
\newblock ``Practical quantum error mitigation for near-future applications''.
\newblock \href{https://dx.doi.org/10.1103/PhysRevX.8.031027}{Phys. Rev. X {\bf
  8}, 031027}~(2018).

\bibitem{mcdonoughAutomatedQuantumError2022}
Benjamin McDonough, Andrea Mari, Nathan Shammah, Nathaniel~T. Stemen, Misty
  Wahl, William~J. Zeng, and Peter~P. Orth.
\newblock ``Automated quantum error mitigation based on probabilistic error
  reduction''.
\newblock In 2022 IEEE/ACM Third International Workshop on Quantum Computing
  Software (QCS).
\newblock \href{https://dx.doi.org/10.1109/QCS56647.2022.00015}{Pages 83--93}.
\newblock ~(2022).

\bibitem{van_den_Berg21}
Ewout van~den Berg, Zlatko~K. Minev, Abhinav Kandala, and Kristan Temme.
\newblock ``Probabilistic error cancellation with sparse pauli–lindblad
  models on noisy quantum processors''.
\newblock \href{https://dx.doi.org/10.1038/s41567-023-02042-2}{Nature Physics
  {\bf 19}, 1116–1121}~(2023).

\bibitem{Mitarai19}
Kosuke Mitarai and Keisuke Fujii.
\newblock ``Methodology for replacing indirect measurements with direct
  measurements''.
\newblock \href{https://dx.doi.org/10.1103/PhysRevResearch.1.013006}{Phys. Rev.
  Research {\bf 1}, 013006}~(2019).

\bibitem{Bernien17}
Hannes Bernien, Sylvain Schwartz, Alexander Keesling, Harry Levine, Ahmed
  Omran, Hannes Pichler, Soonwon Choi, Alexander~S Zibrov, Manuel Endres,
  Markus Greiner, et~al.
\newblock ``Probing many-body dynamics on a 51-atom quantum simulator''.
\newblock \href{https://dx.doi.org/10.1038/nature24622}{Nature {\bf 551},
  579}~(2017).

\bibitem{Bluvstein21}
D.~Bluvstein, A.~Omran, H.~Levine, A.~Keesling, G.~Semeghini, S.~Ebadi, T.~T.
  Wang, A.~A. Michailidis, N.~Maskara, W.~W. Ho, and et~al.
\newblock ``Controlling quantum many-body dynamics in driven {Rydberg} atom
  arrays''.
\newblock \href{https://dx.doi.org/10.1126/science.abg2530}{Science {\bf 371},
  1355}~(2021).

\bibitem{Fendley04}
Paul Fendley, K.~Sengupta, and Subir Sachdev.
\newblock ``Competing density-wave orders in a one-dimensional hard-boson
  model''.
\newblock \href{https://dx.doi.org/10.1103/physrevb.69.075106}{Phys.~Rev.~B
  {\bf 69}, 075106}~(2004).

\bibitem{Turner18a}
C.~J. Turner, A.~A. Michailidis, D.~A. Abanin, M.~Serbyn, and Z.~Papi\'c.
\newblock ``Weak ergodicity breaking from quantum many-body scars''.
\newblock
  \href{https://dx.doi.org/https://doi.org/10.1038/s41567-018-0137-5}{Nature
  Physics {\bf 14}, 579}~(2018).

\bibitem{Turner18b}
C.~J. Turner, A.~A. Michailidis, D.~A. Abanin, M.~Serbyn, and
  Z.~Papi\ifmmode~\acute{c}\else \'{c}\fi{}.
\newblock ``Quantum scarred eigenstates in a rydberg atom chain: Entanglement,
  breakdown of thermalization, and stability to perturbations''.
\newblock \href{https://dx.doi.org/10.1103/PhysRevB.98.155134}{Phys. Rev. B
  {\bf 98}, 155134}~(2018).

\bibitem{joshi_observing_2022}
M.~K. Joshi, F.~Kranzl, A.~Schuckert, I.~Lovas, C.~Maier, R.~Blatt, M.~Knap,
  and C.~F. Roos.
\newblock ``Observing emergent hydrodynamics in a long-range quantum magnet''.
\newblock \href{https://dx.doi.org/10.1126/science.abk2400}{Science {\bf 376},
  720--724}~(2022).

\bibitem{Goldstein06}
Sheldon Goldstein, Joel~L. Lebowitz, Roderich Tumulka, and Nino Zangh\`{\i}.
\newblock ``Canonical typicality''.
\newblock \href{https://dx.doi.org/10.1103/PhysRevLett.96.050403}{Phys. Rev.
  Lett. {\bf 96}, 050403}~(2006).

\bibitem{Popescu06}
Sandu Popescu, Anthony~J. Short, and Andreas Winter.
\newblock ``Entanglement and the foundations of statistical mechanics''.
\newblock \href{https://dx.doi.org/https://doi.org/10.1038/nphys444}{Nature
  Physics {\bf 2}, 043027}~(2006).

\bibitem{gemmer_distribution_2003}
J.~Gemmer and G.~Mahler.
\newblock ``Distribution of local entropy in the {Hilbert} space of bi-partite
  quantum systems: origin of {Jaynes}' principle''.
\newblock \href{https://dx.doi.org/10.1140/epjb/e2003-00029-3}{The European
  Physical Journal B - Condensed Matter and Complex Systems {\bf 31},
  249--257}~(2003).

\bibitem{NielsenChuang}
Michael~A. Nielsen and Isaac~L. Chuang.
\newblock ``Quantum computation and quantum information''.
\newblock \href{https://dx.doi.org/10.1017/cbo9780511976667}{Cambridge
  University Press}. ~(2009).

\bibitem{Brando2016}
Fernando G. S.~L. Brand{\~{a}}o, Aram~W. Harrow, and Micha{\l} Horodecki.
\newblock ``Local random quantum circuits are approximate polynomial-designs''.
\newblock \href{https://dx.doi.org/10.1007/s00220-016-2706-8}{Communications in
  Mathematical Physics {\bf 346}, 397--434}~(2016).

\bibitem{hunter-jones_unitary_2019}
Nicholas Hunter-Jones.
\newblock ``Unitary designs from statistical mechanics in random quantum
  circuits''~(2019) \href{http://arxiv.org/abs/1905.12053}{arXiv:1905.12053}.

\bibitem{Deutsch91}
J.~M. Deutsch.
\newblock ``Quantum statistical mechanics in a closed system''.
\newblock \href{https://dx.doi.org/10.1103/PhysRevA.43.2046}{Phys. Rev. A {\bf
  43}, 2046--2049}~(1991).

\bibitem{Srednicki94}
Mark Srednicki.
\newblock ``Chaos and quantum thermalization''.
\newblock \href{https://dx.doi.org/10.1103/PhysRevE.50.888}{Phys. Rev. E {\bf
  50}, 888--901}~(1994).

\bibitem{srednicki_approach_1999}
Mark Srednicki.
\newblock ``The approach to thermal equilibrium in quantized chaotic systems''.
\newblock \href{https://dx.doi.org/10.1088/0305-4470/32/7/007}{Journal of
  Physics A: Mathematical and General {\bf 32}, 1163}~(1999).

\bibitem{D'Alessio16}
Luca D'Alessio, Yariv Kafri, Anatoli Polkovnikov, and Marcos Rigol.
\newblock ``From quantum chaos and eigenstate thermalization to statistical
  mechanics and thermodynamics''.
\newblock \href{https://dx.doi.org/10.1080/00018732.2016.1198134}{Adv.~Phys.
  {\bf 65}, 239--362}~(2016).

\bibitem{riddell_concentration_2022}
Jonathon Riddell, Nathan~J. Pagliaroli, and Álvaro M.~Alhambra.
\newblock ``{Concentration of quantum equilibration and an estimate of the
  recurrence time}''.
\newblock \href{https://dx.doi.org/10.21468/SciPostPhys.15.4.165}{SciPost Phys.
  {\bf 15}, 165}~(2023).

\bibitem{Ortiz01}
G.~Ortiz, J.~E. Gubernatis, E.~Knill, and R.~Laflamme.
\newblock ``Quantum algorithms for fermionic simulations''.
\newblock \href{https://dx.doi.org/10.1103/PhysRevA.64.022319}{Phys. Rev. A
  {\bf 64}, 022319}~(2001).

\bibitem{Somma02}
R.~Somma, G.~Ortiz, J.~E. Gubernatis, E.~Knill, and R.~Laflamme.
\newblock ``Simulating physical phenomena by quantum networks''.
\newblock \href{https://dx.doi.org/10.1103/PhysRevA.65.042323}{Phys. Rev. A
  {\bf 65}, 042323}~(2002).

\bibitem{Barenco97}
Adriano Barenco, Andr\'{e} Berthiaume, David Deutsch, Artur Ekert, Richard
  Jozsa, and Chiara Macchiavello.
\newblock ``Stabilization of quantum computations by symmetrization''.
\newblock \href{https://dx.doi.org/10.1137/S0097539796302452}{SIAM Journal on
  Computing {\bf 26}, 1541--1557}~(1997).

\bibitem{Buhrman01}
Harry Buhrman, Richard Cleve, John Watrous, and Ronald de~Wolf.
\newblock ``Quantum fingerprinting''.
\newblock \href{https://dx.doi.org/10.1103/PhysRevLett.87.167902}{Phys. Rev.
  Lett. {\bf 87}, 167902}~(2001).

\bibitem{Trotter59}
H.~F. Trotter.
\newblock ``On the product of semi-groups of operators''.
\newblock
  \href{https://dx.doi.org/10.1090/s0002-9939-1959-0108732-6}{Proceedings of
  the American Mathematical Society {\bf 10}, 545--551}~(1959).

\bibitem{Suzuki76}
Masuo Suzuki.
\newblock ``Generalized trotter{\textquotesingle}s formula and systematic
  approximants of exponential operators and inner derivations with applications
  to many-body problems''.
\newblock \href{https://dx.doi.org/10.1007/bf01609348}{Communications in
  Mathematical Physics {\bf 51}, 183--190}~(1976).

\bibitem{Chen22}
I-Chi Chen, Benjamin Burdick, Yongxin Yao, Peter~P. Orth, and Thomas Iadecola.
\newblock ``Error-mitigated simulation of quantum many-body scars on quantum
  computers with pulse-level control''.
\newblock \href{https://dx.doi.org/10.1103/PhysRevResearch.4.043027}{Phys. Rev.
  Res. {\bf 4}, 043027}~(2022).

\bibitem{Borla19}
Umberto Borla, Ruben Verresen, Fabian Grusdt, and Sergej Moroz.
\newblock ``Confined phases of one-dimensional spinless fermions coupled to
  ${Z}_{2}$ gauge theory''.
\newblock \href{https://dx.doi.org/10.1103/PhysRevLett.124.120503}{Phys. Rev.
  Lett. {\bf 124}, 120503}~(2020).

\bibitem{chen_minimally_2024}
I.-Chi Chen, João~C. Getelina, Klée Pollock, Srimoyee Sen, Yong-Xin Yao, and
  Thomas Iadecola.
\newblock ``Minimally {Entangled} {Typical} {Thermal} {States} for {Classical}
  and {Quantum} {Simulation} of {Gauge} {Theories} at {Finite} {Temperature}
  and {Density}''~(2024)
  \href{http://arxiv.org/abs/2407.11949}{arXiv:2407.11949}.

\bibitem{Tindall23}
Joseph Tindall, Matthew Fishman, E.~Miles Stoudenmire, and Dries Sels.
\newblock ``Efficient tensor network simulation of ibm's eagle kicked ising
  experiment''.
\newblock \href{https://dx.doi.org/10.1103/PRXQuantum.5.010308}{PRX Quantum
  {\bf 5}, 010308}~(2024).

\bibitem{Begusic23}
Tomislav Begušić, Johnnie Gray, and Garnet Kin-Lic Chan.
\newblock ``Fast and converged classical simulations of evidence for the
  utility of quantum computing before fault tolerance''.
\newblock \href{https://dx.doi.org/10.1126/sciadv.adk4321}{Science Advances
  {\bf 10}, eadk4321}~(2024).

\bibitem{Saroni23}
Jason Saroni, Henry Lamm, Peter~P. Orth, and Thomas Iadecola.
\newblock ``Reconstructing thermal quantum quench dynamics from pure states''.
\newblock \href{https://dx.doi.org/10.1103/PhysRevB.108.134301}{Phys. Rev. B
  {\bf 108}, 134301}~(2023).

\bibitem{qite_chan20}
Mario Motta, Chong Sun, Adrian~TK Tan, Matthew~J O’Rourke, Erika Ye, Austin~J
  Minnich, Fernando~GSL Brand{\~a}o, and Garnet Kin-Lic Chan.
\newblock ``Determining eigenstates and thermal states on a quantum computer
  using quantum imaginary time evolution''.
\newblock \href{https://dx.doi.org/10.1038/s41567-019-0704-4}{Nat. Phys. {\bf
  16}, 205--210}~(2020).

\bibitem{VQITE}
Sam McArdle, Tyson Jones, Suguru Endo, Ying Li, Simon~C Benjamin, and Xiao
  Yuan.
\newblock ``Variational ansatz-based quantum simulation of imaginary time
  evolution''.
\newblock \href{https://dx.doi.org/10.1038/s41534-019-0187-2}{npj Quantum Inf.
  {\bf 5}, 75}~(2019).

\bibitem{AVQITE}
Niladri Gomes, Anirban Mukherjee, Feng Zhang, Thomas Iadecola, Cai-Zhuang Wang,
  Kai-Ming Ho, Peter~P Orth, and Yong-Xin Yao.
\newblock ``Adaptive variational quantum imaginary time evolution approach for
  ground state preparation''.
\newblock \href{https://dx.doi.org/10.1002/qute.202100114}{Adv. Quantum
  Technol. {\bf 4}, 2100114}~(2021).

\bibitem{Sun2021QuantumCO}
Shi-Ning Sun, Mario Motta, Ruslan~N Tazhigulov, Adrian~TK Tan, Garnet Kin-Lic
  Chan, and Austin~J Minnich.
\newblock ``Quantum computation of finite-temperature static and dynamical
  properties of spin systems using quantum imaginary time evolution''.
\newblock \href{https://dx.doi.org/10.1103/PRXQuantum.2.010317}{PRX Quantum
  {\bf 2}, 010317}~(2021).

\bibitem{AVQMETTS}
João~C. Getelina, Niladri Gomes, Thomas Iadecola, Peter~P. Orth, and Yong-Xin
  Yao.
\newblock ``{Adaptive variational quantum minimally entangled typical thermal
  states for finite temperature simulations}''.
\newblock \href{https://dx.doi.org/10.21468/SciPostPhys.15.3.102}{SciPost Phys.
  {\bf 15}, 102}~(2023).

\bibitem{riddell_no-resonance_2023}
Jonathon Riddell and Nathan Pagliaroli.
\newblock ``No-resonance conditions, random matrices, and quantum chaotic
  models''~(2023) \href{http://arxiv.org/abs/2307.05417}{arXiv:2307.05417}.

\bibitem{dymarsky_new_2019}
Anatoly Dymarsky and Hong Liu.
\newblock ``New characteristic of quantum many-body chaotic systems''.
\newblock \href{https://dx.doi.org/10.1103/PhysRevE.99.010102}{Physical Review
  E {\bf 99}, 010102}~(2019).

\bibitem{capizzi_hydrodynamics_2024}
Luca Capizzi, Jiaozi Wang, Xiansong Xu, Leonardo Mazza, and Dario Poletti.
\newblock ``Hydrodynamics and the eigenstate thermalization
  hypothesis''~(2024).
\newblock  \href{http://arxiv.org/abs/2405.16975v1}{arXiv:2405.16975v1}.

\bibitem{hartmann_gaussian_2004}
Michael Hartmann, Günter Mahler, and Ortwin Hess.
\newblock ``Gaussian {Quantum} {Fluctuations} in {Interacting} {Many}
  {Particle} {Systems}''.
\newblock \href{https://dx.doi.org/10.1023/B:MATH.0000043321.00896.86}{Letters
  in Mathematical Physics {\bf 68}, 103--112}~(2004).

\bibitem{Joydip12}
Joydip Ghosh, Austin~G. Fowler, and Michael~R. Geller.
\newblock ``Surface code with decoherence: An analysis of three superconducting
  architectures''.
\newblock \href{https://dx.doi.org/10.1103/PhysRevA.86.062318}{Phys. Rev. A
  {\bf 86}, 062318}~(2012).

\bibitem{collins_integration_2006}
Benoît Collins and Piotr Śniady.
\newblock ``Integration with {Respect} to the {Haar} {Measure} on {Unitary},
  {Orthogonal} and {Symplectic} {Group}''.
\newblock \href{https://dx.doi.org/10.1007/s00220-006-1554-3}{Communications in
  Mathematical Physics {\bf 264}, 773--795}~(2006).

\bibitem{rastegin_relations_2012}
Alexey~E. Rastegin.
\newblock ``Relations for {Certain} {Symmetric} {Norms} and {Anti}-norms
  {Before} and {After} {Partial} {Trace}''.
\newblock \href{https://dx.doi.org/10.1007/s10955-012-0569-8}{Journal of
  Statistical Physics {\bf 148}, 1040--1053}~(2012).

\end{thebibliography}

\end{document}